\documentclass[aps,prl,showpacs,twocolumn,superscriptaddress,amsmath,amssymb,floatfix, longbibliography]{revtex4-1}
\usepackage{graphicx}
\usepackage{float}
\usepackage{latexsym,amsmath,amssymb,bm,euscript}
\usepackage{color}
\usepackage{wasysym}
\usepackage{epstopdf}
\usepackage[colorlinks=true,linkcolor=blue,citecolor=blue]{hyperref}
\usepackage{type1cm}
\usepackage{appendix}

\usepackage{color}
\definecolor{darkblue}{rgb}{0.,0.,0.4}
\definecolor{darkred}{rgb}{0.5,0.,0.}
\definecolor{BlueViolet}{RGB}{138,43,226}
\definecolor{SkyBlue}{RGB}{30,144,255}
\definecolor{DarkGreen}{RGB}{0,100,0}

\usepackage[normalem]{ulem}

\renewcommand{\vec}[1]{\bm{#1}}

\begin{document}
\title{Dirac Spin Liquid on the Spin-$1/2$ Triangular Heisenberg Antiferromagnet}
\author{Shijie Hu}
\email{shijiehu@physik.uni-kl.de}
\affiliation{Department of Physics and Research Center Optimas, Technische Universitat Kaiserslautern, 67663 Kaiserslautern, Germany}
\author{W. Zhu} 
\email{zhuwei@westlake.edu.cn} 
\affiliation{Institute of Natural Sciences, Westlake Institute of Advanced Study, \\  
School of Science, Westlake University, Hangzhou, 030024, P. R. China}

\author{Sebastian Eggert}
\affiliation{Department of Physics and Research Center Optimas, Technische Universitat Kaiserslautern, 67663 Kaiserslautern, Germany}

\author{Yin-Chen He}
\email{yinchenhe@perimeterinstitute.ca}
\affiliation{Perimeter Institute for Theoretical Physics, Waterloo, Ontario N2L 2Y5, Canada }

\begin{abstract}
We study the spin liquid candidate  of the spin-$1/2$ $J_1$-$J_2$ Heisenberg antiferromagnet on the triangular lattice by means of density matrix renormalization group (DMRG) simulations.
By applying an external Aharonov-Bohm flux insertion in an infinitely long cylinder, we find unambiguous evidence for gapless $U(1)$ Dirac spin liquid behavior.
The flux insertion overcomes the finite size restriction for energy gaps and clearly shows gapless behavior at the expected wave-vectors. Using  the DMRG transfer matrix, the low-lying excitation spectrum can be extracted, which shows characteristic Dirac cone structures of both spinon-bilinear and monopole excitations.
Finally, we confirm that the entanglement entropy follows the predicted universal response under the flux insertion.

\end{abstract}


\date{\today}
\maketitle

\textit{Introduction.}---Quantum spin liquids (QSLs) are exotic phases of matter which remain disordered 
due to quantum fluctuations, which in turn give rise to
remarkable properties of fundamental importance,
such as fractionalizations, gauge fluctuations, topology, and  unconventional superconductivity \cite{SavaryBalentsReview,ZhouReview, Wen_review, SenthilReview}. 
However, despite of a long-running quest, theoretical and experimentally relevant models for enigmatic QSLs are still limited and rare.

Historically, it has been proposed that geometric
frustrations on the spin-$1/2$ triangular antiferromagnetic Heisenberg model (TAFM) could lead to a spin disordered ground state  \cite{Anderson1973}.
Although the nearest neighbor TAFM turns out to exhibit a $120^\circ$ magnetic order \cite{Sachdev1992,Lhuillier1992,Capriotti1999,WZheng2006,SWhite2007},
the possibility of increasing the frustration by adding next-nearest-neighbor (NNN) interactions has captured much interest in the literature~\cite{Manuel1999,Ryan2013,Ryui2014,Iqbal2016,Bishop2015,Wenjun2015,ZZhu2015,
McCulloch2016,Szasz2018,Lauchli2017,Gong2017,Bauer2017,Yuste2018,Ferrari2019,Jolicoeur1990,Jolicoeur1992} for the $J_1$-$J_2$ TAFM
\begin{equation}\label{eq:ham}
 H=J_1 \sum_{\langle i,\,j\rangle} \mathbf{S}_i \cdot \mathbf{S}_j + J_2 \sum_{\langle\langle i,\,j\rangle\rangle} \mathbf{S}_i \cdot \mathbf{S}_j,
\end{equation} 
where $\langle i,\,j\rangle$ and $\langle\langle i,\,j\rangle\rangle$ respectively denote NN and NNN bonds. 
So far, the general consensus is that  
an intermediate region ($0.07\lesssim J_2/J_1\lesssim 0.15$) without magnetic ordering~\cite{Manuel1999,Ryan2013,Ryui2014,Iqbal2016,Bishop2015,Wenjun2015,ZZhu2015,McCulloch2016,Szasz2018,Lauchli2017,Gong2017,Bauer2017,Yuste2018,Ferrari2019} is
sandwiched between a stripe ordered phase ($J_2/J_1\gtrsim0.15$) \cite{Jolicoeur1990,Jolicoeur1992} 
and a $120^\circ$ magnetically ordered phase ($0.0\le J_2/J_1\lesssim0.07$) \cite{Sachdev1992,Lhuillier1992,Capriotti1999,WZheng2006,SWhite2007}. 
However, the underlying physics and precise nature of this intermediate phase is under an intense debate.
For instance, variational Monte Carlo simulations suggest a gapless $U(1)$ Dirac QSL \cite{Iqbal2016} as candidates for this intermediate phase. 
Density-matrix renormalization group (DMRG) calculations \cite{Wenjun2015,ZZhu2015,McCulloch2016,Szasz2018} 
found an indication of a gapped QSL as the nonmagnetic phase, 
while its internal structure  (e.g. Z$_2$, chiral) has yet to be determined.
In addition, extensive exact diagonalization calculations fail to
find evidence in support of either theory in the accessible system sizes \cite{Lauchli2017}.
Taken as whole, although a possible QSL phase has been identified on TAFM, the exact nature of this intermediate phase remains elusive.

It was shown that other experimental-relevant spin models on the triangular lattice also show spin liquid behavior which is continuously connected to the spin-liquid phase of $J_1$-$J_2$ TAFM model~\cite{Zhu2018,Maksimov2019}.
Thus understanding the underlying physics in the $J_1$-$J_2$ TAFM, will give deep insight into a whole class of new triangular materials, for example, the recent synthesized  Na-based chalcogenides \cite{Baenitz2018,WWZhang2018,Ding2019,Ranjith2019,Bordelon2019}. 
In particular, the spin dynamics of NaYbO$_2$ shows low-energy spectral weight accumulating at the $K$-point of the Brillouin zone \cite{Ding2019}.
So far it is unclear if these findings can be interpreted within spin liquid picture \cite{Song2018,Song2018b}, which demonstrates the need for detailed theoretical predictions.

\begin{figure}[b]
\includegraphics[width=0.99\linewidth]{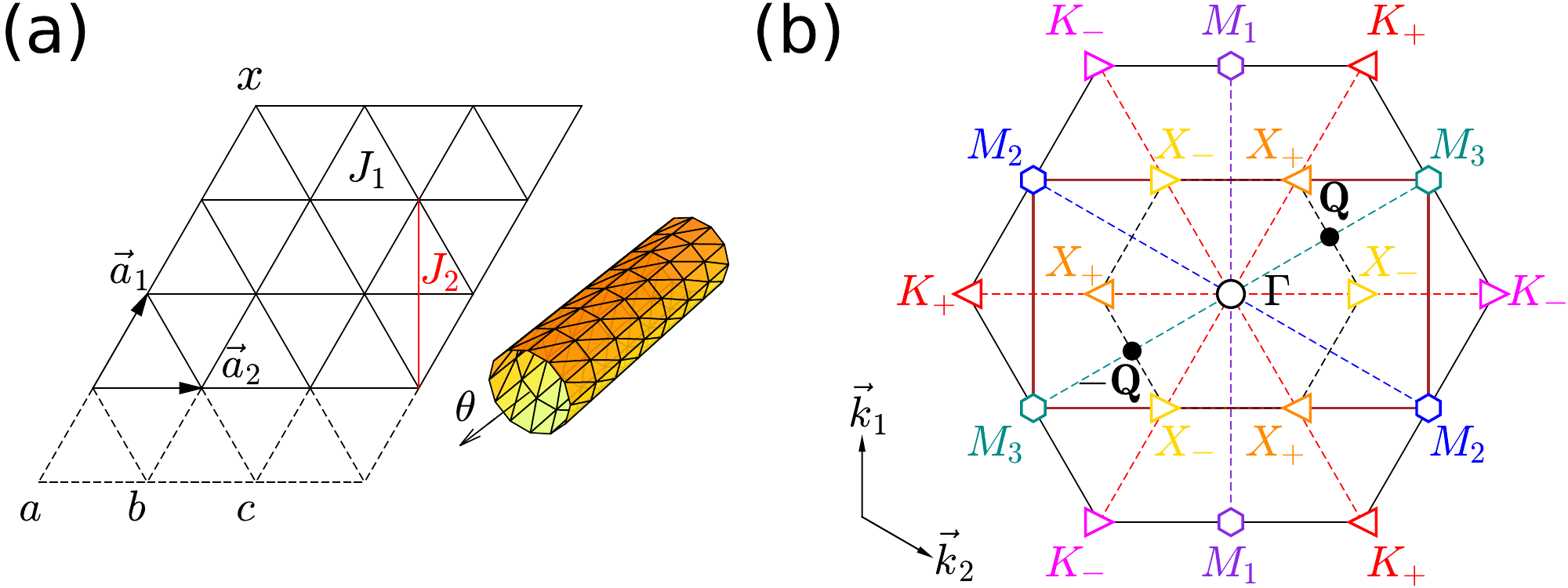}
\caption{Geometry of cylinders in real-space and momentum points in the $1^{\text st}$ Brillouin zone. (a) Three different cylindrical geometries, YC$4$-$0$, YC$4$-$1$ or YC$4$-$2$ \cite{sm}, which correspond to identifying site $x$ with site $a$, $b$, or $c$, respectively.
We also insert an Aharonov-Bohm flux in the hole of the cylinder, which modifies the spin exchange terms in Hamiltonian Eq.~\eqref{eq:ham} across the boundary (labeled by dashed lines).
(b) The black solid line shows the $1^{\text st}$ Brillouin zone of the triangular lattice while the brown rectangle is the magnetic Brillouin zone due to the $\pi$-flux in each unit cell seen by spinons. All characteristic points are labeled by $(k_1,k_2)$ modulo $2\pi$. Two Dirac points (black dots) of spinons are located at $\mathbf{Q}= (\pi/2, \pi/2)$ and $-\mathbf Q$ (details see \cite{sm}). 
The gauge invariant excitations, e.g. fermions bilinears and monopoles, are locating at high symmetric points, including $\Gamma=(0,0)$; $M_1=(\pm\pi, 0)$ (violet $\varhexagon$), $M_2 = (0,\pm\pi)$ (blue $\varhexagon$), $M_3 = \pm(\pi,\pi)$ (dark-cyan $\varhexagon$), $K_{+}=(-2\pi/3,2\pi/3)$ (red $\triangleleft$), $K_{-}=(2\pi/3,-2\pi/3)$ (magenta $\triangleright$); $X_{+}=(-\pi/3,\pi/3)$ (orange $\triangleleft$), $X_{-}=(\pi/3,-\pi/3)$ (gold $\triangleright$).
} \label{fig:geometry}
\end{figure}

In this paper, we unveil the QSL nature of the triangular $J_1$-$J_2$ model by using  large-scale DMRG simulations armed with recently developed state-of-the-art transfer matrix analysis~\cite{He2017,Zauner2014}. 
We find smoking-gun signatures of the $U(1)$ Dirac QSL (DSL), which consistently appear in 16 different geometries and/or system sizes (see Fig.~\ref{fig:geometry}~(a) for details). 
These signatures include: 1) momentum-dependent ``excitation spectra", extracted from the DMRG transfer matrix~\cite{Zauner2014,He2017}, which reveals gapless modes of the Dirac spin liquid showing recently predicted  behavior of both fermion bilinear excitations as well as intricate monopoles~\cite{Song2018,Song2018b}; 
2) strong dependence of the energy gap on twisted boundary conditions~\cite{He2017};
3) universal entanglement entropy response under flux insertion~\cite{He2018}.
These evidences unambiguously show that the intermediate phase in TAFM is a gapless $U(1)$ DSL.

\textit{Properties of $U(1)$ DSL.}---Let us begin with a brief review of properties of the $U(1)$ DSL on the triangular lattice \cite{Zhou_triangle, Iqbal2016, Song2018, Song2018b}.
We begin with rewriting spin operator in terms of fractional fermionic spinons $\vec f=(f_\uparrow, f_\downarrow)^T$, 
$\vec S= \vec f^\dag \vec \sigma \vec f$, where the partons $\vec f$ are coupled to a $U(1)$ dynamic gauge field due to the $U(1)$ redundancy. 
The $U(1)$ DSL can then be realized by putting spinons in a staggered $\pi$ flux mean-field ansatz, whose band structure will have two Dirac cones located at the $\pm Q$ points (valley, black dots in Fig.~\ref{fig:geometry}~(b)) of the Brillouin zone~\cite{Iqbal2016,sm}.
The low energy physics of the $U(1)$ DSL is captured by  $N_f=4$ QED$_3$, namely there are four Dirac fermions (two from spins $\uparrow$/$\downarrow$ and two from valleys) coupled to a dynamic $U(1)$ gauge field.
This $N_f=4$ QED$_3$ theory may flow into a $2$+$1$D conformal field theory (CFT) in the infrared, therefore the $U(1)$ DSL is a critical/conformal phase~\cite{Hermele2004,hermele2005algebraic, Hermele2008}, which is a close analog to the familiar spin-$1/2$ Heisenberg chain in 1+1D~\cite{book,affleck,affleck2}.
One effective way to detect the $U(1)$ DSL is to measure its gapless modes.
It has been shown that the $U(1)$ DSL has two types of fundamental gapless modes, namely fermion (spinon) bilinears and monopoles (of the $U(1)$ gauge field)~\cite{Song2018,Song2018b,Hermele2008,Alicea2008}.
The fermion bilinears are ``particle-hole" excitations of four Dirac fermions, while monopoles are instantons of the $U(1)$ gauge field.
It is worth emphasizing that both fermion bilinears and monopoles are gauge invariant, which correspond to local operators such as spin $\vec S$, dimer operators $\vec S_i \cdot \vec S_j$, etc.
Moreover, these critical operators have distinct quantum numbers (spins, momentum, angular momentum, etc.), enabling us to detect them directly. 

There are in total $16$ fermion bilinears~\cite{hermele2005algebraic}, which can be grouped into $1\oplus 15$, namely $SU(4)$ singlet and adjoint. They are distributed at different momenta in the $1^{\text st}$ Brillouin zone (BZ) of the triangular lattice (seen in Fig.~\ref{fig:geometry}~(b)).
The singlet bilinear is a spin singlet with zero momentum ($\Gamma$ point, black $\Circle$).
The $15$ adjoint bilinears can be classified into three types~\cite{Song2018}: Type {\bf B1)} $3$ time-reversal-even spin singlets with momenta located at three $M$ points of BZ: $M_1$ (violet $\varhexagon$), $M_2$ (blue $\varhexagon$) and $M_3$ (dark-cyan $\varhexagon$). 
Type  {\bf B2)} $3$ time-reversal-even spin triplets with zero momentum located at the $\Gamma$ points (black $\Circle$) of BZ. 
Type {\bf B3)} $9$ time-reversal-odd spin triplets with momenta located at three $M$ points ($\varhexagon$).
The quantum numbers of the monopoles in the $U(1)$ DSL 
remained elusive for decades until recently solved in Ref.~\cite{Song2018,Song2018b}.
There are $6$ monopole operators (which are complex) of two types: Type {\bf M1)} $3$ time-reversal-odd spin-triplets with momenta located at $K_{\pm}$ 
(red $\triangleleft$ and magenta $\triangleright$).
Type {\bf M2)} $3$ time-reversal-even spin-singlets with momenta located at the $X_{\pm}$ points (orange $\triangleleft$ and gold $\triangleright$).
Physically, the condensation of spin-triplet monopoles will give the familiar $120^\circ$ non-colinear magnetic order, while the condensation of spin-singlet monopoles leads to valence bond solid such as $\sqrt{12}\times \sqrt{12}$ state~\cite{Song2018}.
Later we will show that  signatures of both the fermion bilinear and monopole operators have been measured in our DMRG simulations.

\textit{Method.}---We use infinite-DMRG~\cite{White1992,White1993,McCulloch2008} to simulate the $J_1$-$J_2$ TAFM wrapped on infinitely long cylinders.
The evidence for DSL excitations is based on two major improvements in this study.
First, we study different types of cylindrical geometries which correspond to different ways of wrapping a cylinder. 
As shown in Fig.~\ref{fig:geometry}(a), we define the YC$L_{y}$-$n$ cylinders by identifying the site $\vec r$ with the site $\vec r + L_{y} \vec{a}_1- n\vec{a}_2$ \cite{sm}.
For instance, the notation YC$8$-$1$ denotes a cylinder (C) with circumference of eight lattice spacing and a shift of one column in the y direction (Y) when connected periodically.
Here, $\vec{a}_{1/2}$ are the triangular Bravais lattice primitive vectors, $L_y$ is the ``circumference'' of the cylinder, and $n$ amounts to a shift along the cylindrical direction.
Simulating different geometries not only proves the observed DSL signatures are robust against finite size effect, but also serves as a nontrivial check as DSL on different geometries show qualitatively different behaviors (see Supplemental Materials \cite{sm}).
Second, we carry out a numerical Aharonov-Bohm experiment by inserting flux $\theta$ in the cylinder, see Fig.~\ref{fig:geometry}~(a)).  
This is implemented by twisted boundary conditions, which modifies interactions crossing the boundary by a phase factor, e.g. $S^+_{i} S^-_{j} e^{i\theta} + S^+_{j} S^-_{i} e^{-i\theta}$ with a flux angle $\theta$~\cite{He2014a}.
With the flux insertion, we can fully scan the momentum points in the Brillouin zone on a given geometry and are therefore not limited by finite-size energy gaps.
Furthermore, certain physical quantities in DSL, such as the entanglement entropy, have a nontrivial response under flux insertion~\cite{He2018}.

In the simulation it is important to keep the ground-state  evolving adiabatically under the flux insertion. 
In most cases, the adiabatic flux insertion can be maintained, except 
very close to the Dirac cone (large flux $\theta$), where accurate infinite-DMRG simulation becomes very challenging due to the small gap and large entanglement of the state.
Once adiabatic flux insertion fails at large $\theta$, the infinite-DMRG simulation may suddenly collapse  to a competing state in other super-selection sectors of the ground-state~\cite{sm} or a symmetry broken state due to the instability of the gapless state~\cite{He2017}.
We will not present the data of flux $\theta$ for which adiabatic flux insertion fails, as they do not reveal any direct information of spin-liquid ground state at zero flux.

\begin{figure}[b]
\includegraphics[width=0.99\linewidth]{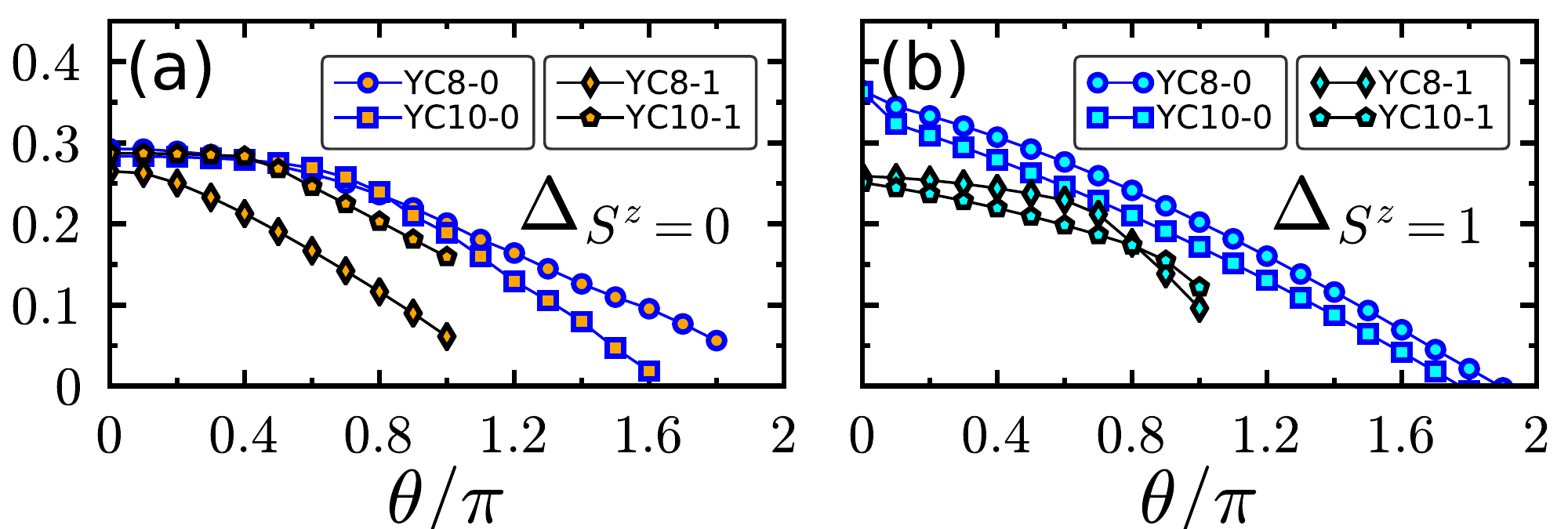}
\caption{Spin excitation gap.
(a) Energy gap $\Delta_{S^z=0}$ and (b) $\Delta_{S^z=1}$ as a function of the inserted flux $\theta$ at $J_2/J_1=0.12$ for YC$8$-$0$  (blue $\Circle$), YC$10$-$0$ (blue {\scriptsize $\square$}), YC$8$-$1$ (black {\scriptsize $\diamondsuit$}) and YC$10$-$1$ (black $\pentagon$) cylinder geometries. 
The data is collected using DMRG bond dimension $m=4096$ for YC$8$/$10$-$0$ and $6144$ for YC$8$/$10$-$1$. 
(Details please see the Suppl. Mat.\cite{sm})
} \label{fig:gap}
\end{figure}

\textit{Excitation gap.}---Previous DMRG studies have found a considerably large spin gap in the $J_1$-$J_2$ TAFM \cite{ZZhu2015}.
However, this is not sufficient to exclude a DSL since 
on a cylinder the momentum is discrete, so the gapless Dirac point may be missed.
The flux insertion, which effectively changes the quantized momentum of spinons, can make spinons hit Dirac point at specific values of flux $\theta$~\cite{He2017}.
By carefully studying the DSL ansatz incorporating the effect of emergent gauge fields \cite{sm}, we find that DSL on different cylinder geometries YC$L_y$-$n$ have distinct $\theta$ dependence.
If both $L_y$ and $n$ are even, spinons are gapless when $\theta=2\pi$ (Since spinons are fractional particles, the flux insertion has $4\pi$ periodicity).
For all other three cases, spinons  will be gapless at $\theta=\pi$ or $3\pi$.

Figure~\ref{fig:gap} shows the energy gap as a function of  flux $\theta$. 
Although the gap is large at $\theta=0$  \cite{note1}, we find it significantly decreases as $\theta$ increases. 
The sensitivity of the energy gap is an indication of the gapless DSL: 1) for a gapped spin liquid the spin gap should have a small dependence (exponentially in $L_y$) on the flux; 2) finite flux drags the momentum lines toward the Dirac points, thus the gap monotonically reduces. 
 Due to the small gap when Dirac points are approached, we are not able to maintain the adiabatic flux insertion when $\theta\sim 1.5\pi$ for YC$2n$-$2m$ cylinder, and $\theta\sim \pi$ for all other cylinders. 
There are also truncation effects from
the finite bond dimensions $m$ in infinite-DMRG, which 
may explain that the YC$10$-$1$ gap appears larger than the YC$8$-$1$ gap in Fig.~\ref{fig:spec}.  
We discuss results for different $m$ in the Supplemental Material \cite{sm}.
We also remark that the gap we measured may come from monopoles (rather than spinons), whose finite size effect is more subtle to analyze.
The important message to take is, in all cases the gap systematically decreases as a
function of $\theta$, and it is consistent with the theoretical expectation that the finite-size
gap of spinons vanishes
at i) $\theta=\pi$ for YC$8$-$1$ and YC$10$-$1$, ii) $\theta=2\pi$ for YC$8$-$0$ and YC$10$-$0$.

\begin{figure}[t]
  \includegraphics[width=0.99\linewidth]{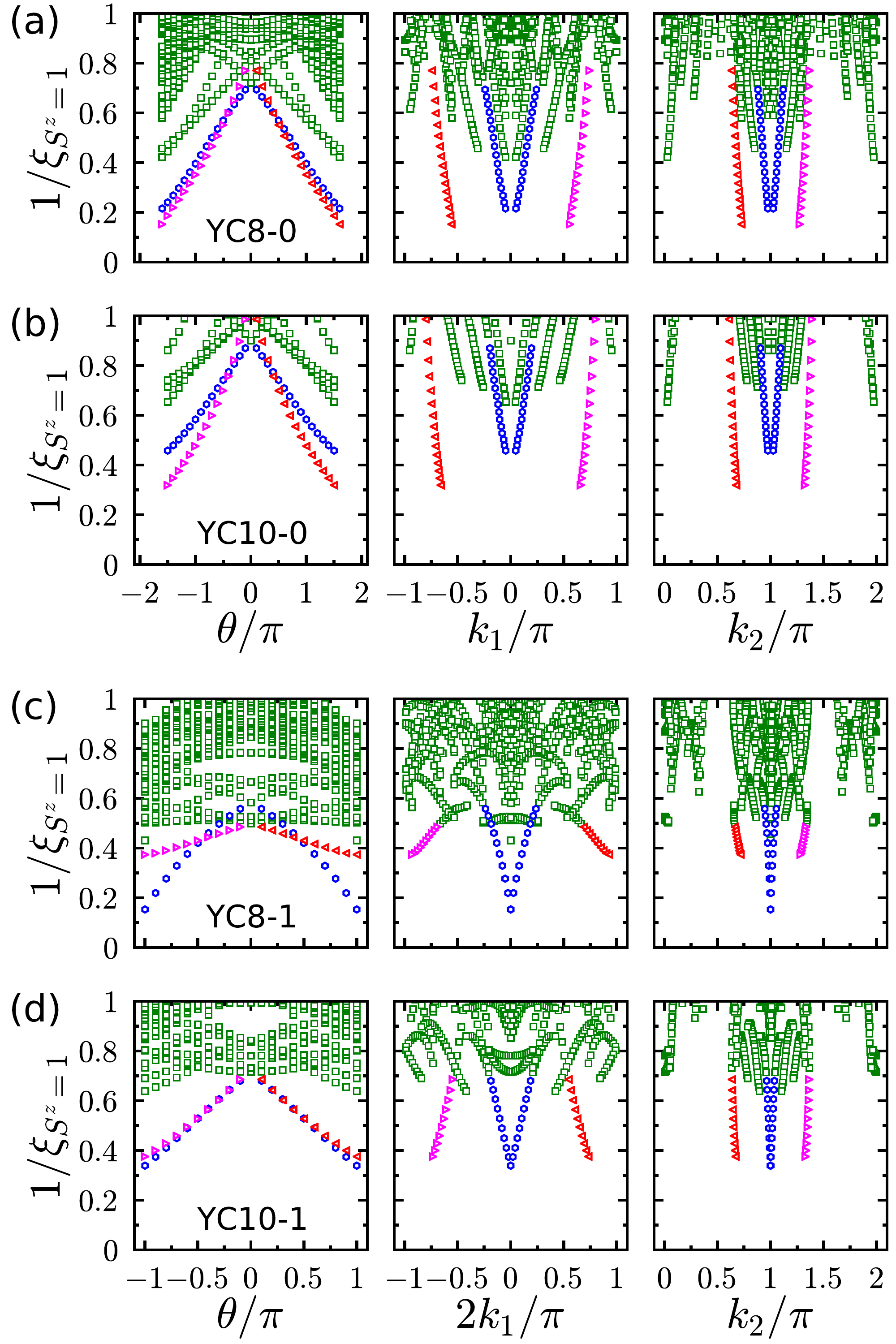}
  \caption{Correlation length spectrum.
Inverse correlation length $1/\xi_{S^z=1}$ as a function of the flux $\theta$ (left column), momentum $k_1$ (middle column) and momentum $k_2$ (right column) for the cylinder (a) YC$8$-$0$, (b) YC$10$-$0$, with $m=6144$ and (c) YC$8$-$1$, (d) YC$10$-$1$ with $m=12288$. 
The lowest-lying excitations contain spinon-pair at $M$ points and monopole excitations at $K_{\pm}$ point, the former is denoted by blue $\varhexagon$ while the latter is denoted by red $\triangleleft$ and magenta $\triangleright$.
It is an artifact of finite bond dimension that the correlation length is not diverging at the Dirac point, and it becomes more severe for the larger system sizes (see Supplemental Material~\cite{sm} for more discussion).
} 
\label{fig:spec}
\end{figure}

\textit{Correlation-length Spectrum.}---While the energy gap is an important indication, the Dirac cone structure of the 
energy-momentum resolved spectrum will be a much 
stronger evidence of a DSL.  
So far the study of a large number of excited states has been very challenging, but fortunately recent seminal works \cite{Zauner2014,He2017} have uncovered a relationship between the energy spectrum and the spectrum of the transfer matrix in tensor-network formulation, which opens a window to the current problem.

The essence of this technique simply relies on a familiar fact: the information of excitations is encoded in the ground state, which can be decoded by measuring correlation functions 
of various operators.
In infinite-DMRG simulations, the information of correlation functions of all operators 
can be straight-forwardly obtained through the 
eigenvalues of the transfer matrix~\cite{SCHOLLWOCK2011}. 
Each eigenvalue takes the form $\lambda =e^{{i k}-1/\xi}$, where  $\xi$ is 
the corresponding correlation length and 
$k$ is the momentum along the infinite-DMRG direction. The momentum around the cylinder can also be calculated from a revised transfer matrix \cite{He2017}.
The correlation lengths $\xi$ set an upper bound for excitation gaps $\Delta$ (up to a  non-universal factor), and for a Lorentz invariant systems it holds that $\Delta \propto 1/\xi$.
One can make this statement precise by an exact mapping from the infinite-DMRG transfer matrix to the partition function in the Euclidean path integral~\cite{Zauner2014}. 
In other words, if Lorentz (space-time rotation) symmetry is emergent in the system, the correlation-length spectrum precisely corresponds to the excitation spectrum of the Hamiltonian.

Fig.~\ref{fig:spec} shows the $S^z=1$ correlation-length spectrum of the $J_1$-$J_2$ TAFM at  $J_2/J_1=0.12$.
The left column shows the spectrum as a function of flux $\theta$. Since $\theta$ effectively changes the quantization of the momenta, we can then obtain the full dispersion relation as a function of $k_1$ and $k_2$ in the two right columns.
For the cylinders YC$8$/$10$-$0$ in Fig.~\ref{fig:spec}~(a)/(b) the Dirac cones at the $M_2$ point $(k_1, k_2)=(0,\pi)$ can clearly be identified (blue $\varhexagon$), corresponding to fermion bilinear excitations of type B3 discussed above.  
In addition there are low lying monopole excitations close to the $K_\pm$ points $(k_1, k_2)=(-2\pi/3,2\pi/3)$ and $(2\pi/3,-2\pi/3)$ (red $\triangleleft$ and magenta $\triangleright$ respectively) of type M1.
For the cylinders YC$8$/$10$-$1$ in Fig.~\ref{fig:spec}~(c)/(d), 
we again find low lying excitations at the $M$ point $(2k_1, k_2)=(0, \pi)$ 
(blue $\varhexagon$) and $K$-points $(2k_1, k_2)=(2\pi/3,2\pi/3)$, $(-2\pi/3,-2\pi/3)$ (red $\triangleleft$ and magenta $\triangleright$).
These observation of low lying excitations of fermion bilinear and monopole operators is clear evidence for a $U(1)$ DSL.  
We note that the lattice rotation symmetry $C_6$ is broken on the cylinder geometry, so its corresponding degeneracy is naturally split.
In the Supplemental Material a total of 16 different cylinder geometries are analyzed, which consistently show the predicted $U(1)$ DSL excitations \cite{sm}.

\textit{Entanglement Entropy.}---Gapless spin liquids have nontrivial  long-ranged quantum entanglement, in contrast to Landau ordered phases. 
We therefore also consider the bipartite entanglement entropy, ${\cal S}=-{\rm Tr}_{\rm sys} (\rho_{\rm sys} \ln \rho_{\rm sys})$, where the reduced density matrix $\rho_{\rm sys}={\rm Tr}_{\rm env} (|\Psi\rangle \langle \Psi|)$ for the half-cylinder ``system" is constructed by ground-state wave function $|\Psi\rangle$ and traced over the degrees of freedom in the other half-cylinder ``environment."
It was recently proposed that the entanglement entropy of $2$+$1$D CFT may have a universal  response to an external Aharonov-Bohm flux~\cite{Xiao2017}.
In particular, for DSL~\cite{He2018}, we expect
\begin{equation}\label{eq:scaling}
{\cal S}\!=S_0(L_y)-\!B\!\sum_{n=1}^{N_f} \ln \left|2\sin \left[\frac{s}{2}\left(\theta-\theta^c_n\right)\right] \right|,
\end{equation}
where $S_0(L_y)$ represents the area law part of entropy and $B$ is a prefactor which may or may not be universal.
Other parameters are universal and can be determined by the underlying theory:
$N_f$ accounts for the number of flavors of different Dirac spinons, 
$s=1/2$ is the fractional spin carried by Dirac spinons,  and 
$\theta^c_n$ corresponds to the flux value 
at which the $n^{\textrm{th}}$ Dirac spinon  becomes gapless.
This scaling function (Eq.~\eqref{eq:scaling}) has been successful applied to  
identify the emergent DSL of the kagome antiferromagnet~\cite{He2018}. 

\begin{figure}[t]
  \includegraphics[width=0.99\linewidth]{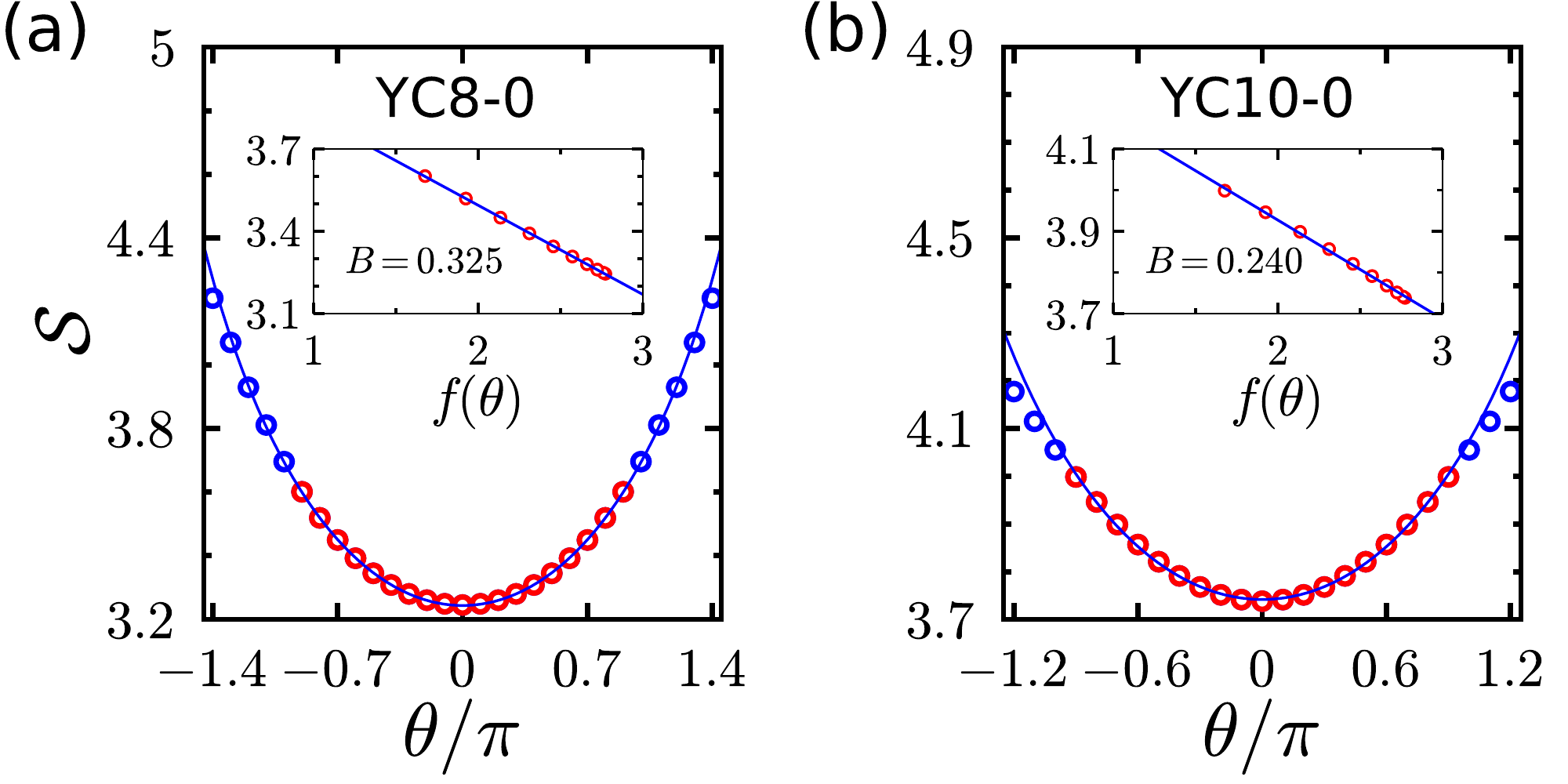}
  \caption{Scaling behavior of the entanglement entropy.
Entanglement entropy $\cal S$ (blue circles) as a function of 
flux $\theta$ for (a) YC$8$-$0$ ($m=12288$) and (b) YC$10$-$0$ ($m=8192$) cylinder geometries.
 Inset: Fit around minima to Eq.~\eqref{eq:scaling}, where
$f(\theta)=\sum_{n=1}^{N_f} \ln \left|2\sin \left[s\left(\theta-\theta^c_n\right) / 2\right] \right|$.
We fit the formula with data labeled by red circles.
} \label{fig:entropy}
\end{figure}

Fig.~\ref{fig:entropy} shows the flux dependence of the 
entanglement entropy $\cal S$ at $J_2/J_1 = 0.12$,
which has a strong dependence on flux $\theta$. This 
is a hallmark of low energy gapless excitations.
In contrast, a fully gapped state would be largely insensitive to $\theta$.
Moreover, as shown in Fig.~\ref{fig:entropy}, the dependence of $\cal S$ on $\theta$ can be 
fitted by the scaling function Eq.~\eqref{eq:scaling} with parameters $N_f=4$, $s=1/2$ and $\theta^c_n=2\pi$ for YC$2n$-$0$ and $\theta^c_n=\pm \pi$ for YC$2n$-$1$. 
This agrees well with our theoretical expectation.

\textit{Summary and Discussion.}---By combining large-scale DMRG simulations 
and recent analytical predictions, we study the intermediate spin liquid phase on the $J_1$-$J_2$ triangular antiferromagnetic Heisenberg model. 
Using flux insertion on different cylinder geometries
we demonstrate that the energy gap of the spin liquid closes, and more importantly we find the low energy excitations of fermion bilinears and monopoles of Dirac spin liquid.
The simultaneous appearance of fermion bilinears and monopoles is in favor of a Dirac spin liquid scenario, as opposed to the scenario of proximity to an ordered phase. Moreover, the entanglement entropy response under flux insertion agrees with a universal scaling law of the Dirac spin liquid.
These findings strongly suggest that 
the intermediate phase of the $J_1$-$J_2$ TAFM is a gapless Dirac spin liquid.

\begin{acknowledgments}
We thank Chong Wang for fruitful discussion.
W.~Z. is supported by start-up funding from Westlake University, and project 11974288 under NSFC.
S.~J.~H. and S.~E. is supported by the German Research Foundation (DFG) via the Collaborative Research Center SFB/TR185 (OSCAR),
 Research at Perimeter Institute (Y.~C.~H.) is supported by the Government of Canada through the Department of Innovation, Science and Economic Development Canada and by the Province of Ontario through the Ministry of Research, Innovation and Science.
Especially, we gratefully acknowledge the Gauss Centre for Supercomputing e.V. (www.gauss-centre.eu) for funding this project by providing computing time through the John von Neumann Institute for Computing (NIC) on the GCS Supercomputer JUWELS at J\"{u}lich Supercomputing Centre (JSC).
\end{acknowledgments}

\bibliography{spin_liquid}

\clearpage

\appendix

\widetext
\begin{center}
	\textbf{\large Supplementary Material for: Dirac Spin Liquid on the Spin-$1/2$ Triangular Heisenberg Antiferromagnet}
\end{center}

\vspace{8mm}

In the Supplemental Material we present details related to mean-field theory of the Dirac spin liquid, the numerical procedure, and more supporting data. First of all, we restate the low energy theory for $N_f=4$ QED$_3$ of the $U(1)$ Dirac spin liquid on the triangular lattice as well as its finite size effect on different geometries.
Secondly, we discuss the numerical observations of even/odd sector states and weak nematic ordering of the groundstate.
At last, we show additional data of transfer-matrix spectrum, entanglement entropy and gaps for total simulated $16$ different geometries and/or system sizes.

\vspace{8mm}

\renewcommand\thefigure{\thesection S\arabic{figure}}

\setcounter{figure}{0} 

\twocolumngrid

\section{$U(1)$ Dirac spin liquid on 
the triangular lattice}
\begin{figure}[b]
    \centering
    \includegraphics[width=0.9\linewidth]{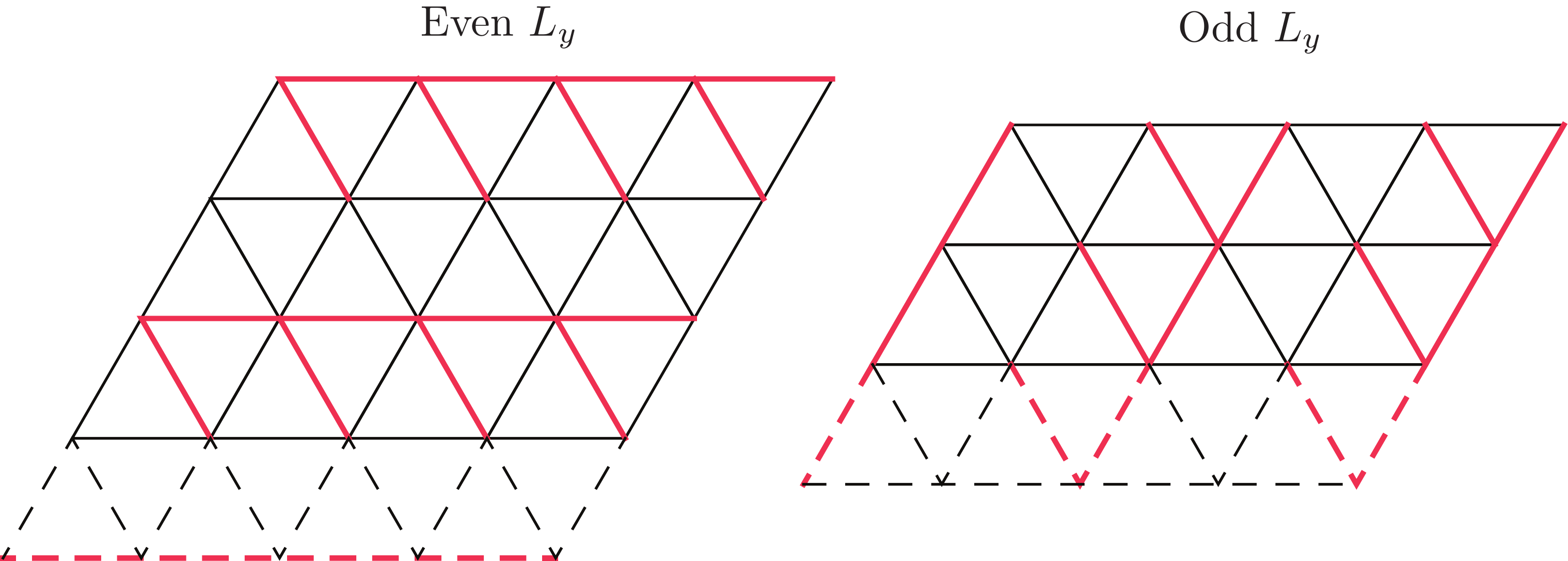}
    \caption{$U(1)$ Dirac spin liquid ansatz on the cylindrical geometry with even/odd width $L_y$. 
    We have hopping amplitude $-1$ on the red bond, $1$ on the black bond.}
    \label{fig:even_odd}
\end{figure}

We fractionalize spin operators into fermionic partons, $\vec S=\vec f^\dag \vec \sigma f$, $\vec f^\dag = (f_\uparrow^\dag, f_\downarrow^\dag)$.
The $U(1)$ Dirac spin liquid (DSL) can be described by a mean-field ansatz, 
\begin{equation}
H_{MF} = -\sum_{\langle ij\rangle} \sum_{\sigma=\uparrow/\downarrow} [(-1)^{s_{ij}} f_{i, \sigma}^\dag f_{j, \sigma} +h.c.].
\end{equation}
$(-1)^{s_{ij}}$ is chosen to give a $\pi$/$0$ flux on the up/down triangles.
One can find a band structure of two Dirac cones in the $1^{\text st}$ Brillouin zone.
Therefore, the low-energy theory of this state is described by $N_f=4$ QED$_3$.

Once we put $U(1)$ DSL on a finite-width and infinite-length cylinder, there are several subtle finite size effects, which can be understood in the mean-field theory.
First, the cylinder may trap an extra flux $\phi=0$ or $\pi$ from the emergent $U(1)$ gauge field~\cite{He2017}.
The value of $\phi$ will be energetically determined.
Second, we note that there is an even-odd effect for the mean-field ansatz.
As shown in Fig.~\ref{fig:even_odd}, cylinders with even and odd circumference $L_y$ should have different gauge configurations.
Next, we solve band structure with an additional phase $e^{i\sigma\theta/2+i\phi}$ for the hoppings across the boundary in axis $\vec{a}_1$,
where $\theta$ is the external Aharonov-Bohm flux inserted in the cylinder and the factor $\sigma/2$ is due to that $f_{\uparrow/\downarrow}$ carries $\pm 1/2$ spin.
So on the YC$L_y$-$n$ cylinders, the properties of the band structure can be classified according to the even or oddness of $L_y$ and $n$.
If both $L_y$ and $n$ are even, $\phi=\pi$ will be energetically favored, and the fermions become gapless when $\theta=2\pi$.
For the other cases, the system behaves qualitatively similar no matter $\phi=0$ or $\pi$, and
the fermions will become gapless at $\theta=\pi$ or $3\pi$, where only two (instead of four in the thermodynamic limit) Dirac fermions become gapless.

\section{Transfer matrix technique}

\begin{figure}[t]
\includegraphics[width=0.95\linewidth]{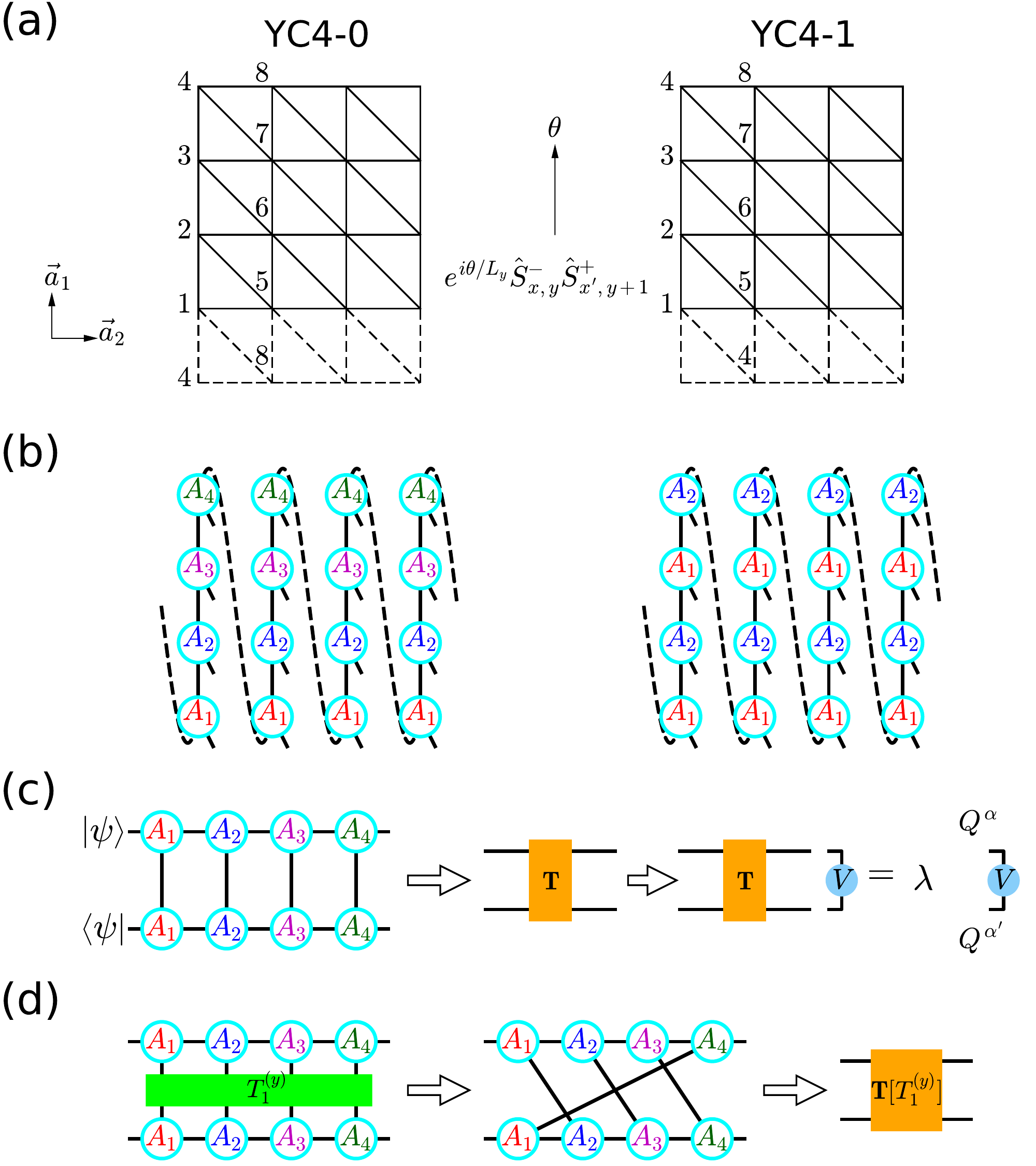}
\caption{(a) Equivalent square-lattice geometry for cylinders YC$4$-$0$ and YC$4$-$1$ on the triangular lattice. Bravais lattice primitive vectors ${\mathbf a}_{1/2}$ are indeed x/y axes. Inserted flux $\theta$ leads to a phase $e^{i\theta / L_y}$ in front of the spin-flipping term ${\hat S}^{-}_{x,y} {\hat S}^{+}_{x',y+1}$. (b) MPS representations for cylinders YC$4$-$0$ and YC$4$-$1$. (c) Pure transfer-matrix $\mathbf T$ of a smallest repeating unit cell. Its eigenvalues are determined by the quantum number discrepancy $q=Q^{\alpha}-Q^{\alpha'}$. (d) The mixed transfer matrix ${\mathbf T}[T^{(y)}_{1}]$ within a translation operation $T^{(y)}_{1}$ under one site along the cylinder circumference to calculate the momentum $k_1$ of each Schmidt basis.
\label{fig:tmtech}
} 
\end{figure}

In this section, we provide more details of the transfer matrix technique.
In infinite-DMRG, we wrap a 2D lattice on a thin cylinder with a finite circumference $L_y$ but an infinite length $L_x$ and use the snake-chain matrix product state (MPS) to cover the cylinder as shown in Fig.~\ref{fig:tmtech}~(b) (for simplicity we view triangular lattice as a square lattice with a diagonal bond).
Generically the snake-chain MPS geometry breaks the lattice translation symmetry along the $a_1$ direction, so we need to use $2n$ distinct matrices $A_{l}$ ($l=1$, $\cdots$, $2n$) in the MPS of  YC$2n$-$0$ cylinders (see Fig.~\ref{fig:tmtech}~(b), left panel), while $4n+2$ matrices for YC$(2n+1)$-$0$ cylinder.
For a special geometry, i.e. YC$L_y$-$1$, MPS recover two-site repeating structure, namely $A_{1}$-$A_{2}$, independent if $L_y$ is even or odd (see Fig.~\ref{fig:tmtech}~(b), right panel).
On the other hand, the MPS is translation invariant and repeating along the direction $a_{2}$. Then, one can use the smallest repeating unit cell to define the transfer matrix (TM), as shown in Fig.~\ref{fig:tmtech}~(c).
The eigenvalues $\lambda_{j}= e^{ ik_{j}-\xi^{-1}_{j}}$ of TM--${\mathbf T}$ contain information of the correlation functions of all operators, which are further related to excitations of the system~\cite{Zauner2014}.
Physically, each eigenvalue corresponds to one excitation mode of the system: $\xi_{j}$ gives the correlation length (or equivalently the inverse of gap), while $k_{j}$ gives the momentum along the infinite direction of the cylinder.

We can further extract the conserved quantum number of each excitation mode from TM.
We note that the Schmidt basis (virtual index) of MPS has a well defined conserved quantum number ($Q^\alpha$), hence each eigenvector of TM--$\mathbf T$ shall have a quantum number $Q=Q^\alpha-Q^{\alpha'}$ (Fig.~\ref{fig:tmtech}(c)).
From this calculation we can get the correlation-length spectrum of different $S^z$ sectors.
  Below we discuss the method of calculating momentum $(k_1,k_2)$ of the correlation length spectrum.
  
 For the YC$L_y$-0 geometry, $k_2$ is nothing but $k$ from the eigenvalues ($e^{ik-\xi^{-1}}$) of TM--$\mathbf T$,
 $k_1$ on the other hand requires a bit more work.
 Due to the snake covering the MPS does not have translational invariance along the compactified direction ($\bf a_1$).
However, the Hamiltonian still has translational invariance along $\bf a_1$, hence the momentum $k_1$ is a conserved quantum number.
In other words, $k_1$ is similar to other conserved quantum number (e.g. $S^z$) even though it is not encoded in MPS explicitly.
To extract $k_1$ of each mode one needs to first obtain the conserved momentum $ k_1$ of each Schmidt basis of the MPS using the mixed transfer matrix TM--$\mathbf T[T^{(y)}_{1}]$.  The mixed  TM--$\mathbf T[T^{(y)}_{1}]$ is pictorially defined in Fig.~\ref{fig:tmtech}(d), namely it is defined by translating the MPS by one site along the $\bf a_1$ direction.
Due to the translation invariance (along $\bf a_1$) the dominant eigenvector of ${\mathbf T}[T^{(y)}_{1}]$ should be $\mathbf V_{\alpha, \alpha'}=\delta_{\alpha, \alpha'} e^{i \bf k^\alpha}$, with $\bf k^\alpha$ being the conserved momentum $ k_1$ of each Schmidt basis of the MPS.
At last, each mode has the momentum $k_1=\bf k^\alpha-\bf k^{\alpha'}$, where $\alpha$ and $\alpha'$ are the Schmidt basis of eigenvectors of TM--$\bf T$.
In the simulation we inserted a finite flux $\theta$ in the cylinder, so  the aforementioned momentum is further modified $(k_1+\theta/L_y, k_2)$ to obtain the final correlation spectrum presented in the paper.

For the special geometry YC$L_y$-$1$, the momenta $k_1$ and $k_2$ are intertwined together. 
A benefit is that the MPS with snake-geometry is actually invariant under two-sites translation. 
And the momentum $(k_1, k_2)$  can be obtained directly form the TM's eigenvalues $e^{ik-1/\xi}$,
\begin{eqnarray}
2k_{1} = k + 2\theta/L_{y},\quad k_2 = k L_{y}/2.
\end{eqnarray}
Therefore, for the $-1$ geometry, one can still get $k_{1}$ and $k_{2}$, but $k_{1}$ has a $\pi$ ambiguity.

\section{Even/odd-sector states}

It is well known that a topological ordered state has a number of topological degenerate groundstates once it is placed on a torus or an infinite cylinder (e.g. see~\cite{He2014a}). Physically all topological sectors can be understood as a state with gauge fluxes and/or gauge charge lines (i.e. anyon lines) threaded in the torus or cylinder.
Similarly the $U(1)$ DSL also has different groundstate sectors, which we will call as super-selection sectors instead of topological sectors.
The properties of super-selection sectors of $U(1)$ DSL are not well understood theoretically, and it will not be pursued here either. 
In this section we will try to clarify some confusion regarding the topological/super-selection sectors obtained in DMRG simulations.

For a spin liquid Hamiltonian, DMRG simulations may yield several different ``groundstates" (They are local minima of the Hamiltonian). However, there is no a-priori knowledge that these ``groundstates" are different topological/super-selection sectors of a spin liquid phase.
One has to conduct a thorough study on these ``groundstates". 
For example, for a gapped topological spin liquid one needs to check if the modular matrix calculated from these ``groundstates" agrees with the theoretical expectation for a topologically ordered phase.
Otherwise one cannot exclude the possibility that different ``groundstates" are the groundstates of different competing phases. 

For the $J_1$-$J_2$ triangular spin liquid, one can obtain two ``groundstates" on the YC$2n$-0 geometry. 
Previous works~\cite{Wenjun2015,ZZhu2015,McCulloch2016,Gong2017} call these two ``groundstates" the even and odd sectors, which pictorially corresponds to the number of valence bonds which cross a cut through the system. Their entanglement spectrum is one-fold or two-fold degenerate respectively,  related to whether the $SO(3)$ spin rotation symmetry is realized projectively (two-fold) or not (one-fold).
Numerically, we can get even/odd-sector states in the model without/with a pair of blank sites center-symmetrically located at two edge-columns during infinite-DMRG ``warming-up" steps. 
For three cylindrical geometries (YC$2n$-$(2k+1)$, YC$(2n+1)$-$2k$, YC$(2n+1)$-$(2k+1)$) there is no distinction between the even and odd sector, as these two sectors are simply related by a translation symmetry.
In contrast, on the YC$2n$-$2k$ cylinders the odd sector and even sector are different states, and we mostly focus on the odd sector for this geometry.
The even sector on the YC$2n$-$2k$ cylinders has a smaller gap, which also shows signatures of $U(1)$ DSL in the correlation length spectrum.    

\begin{figure}[b]
\includegraphics[width=0.9\linewidth]{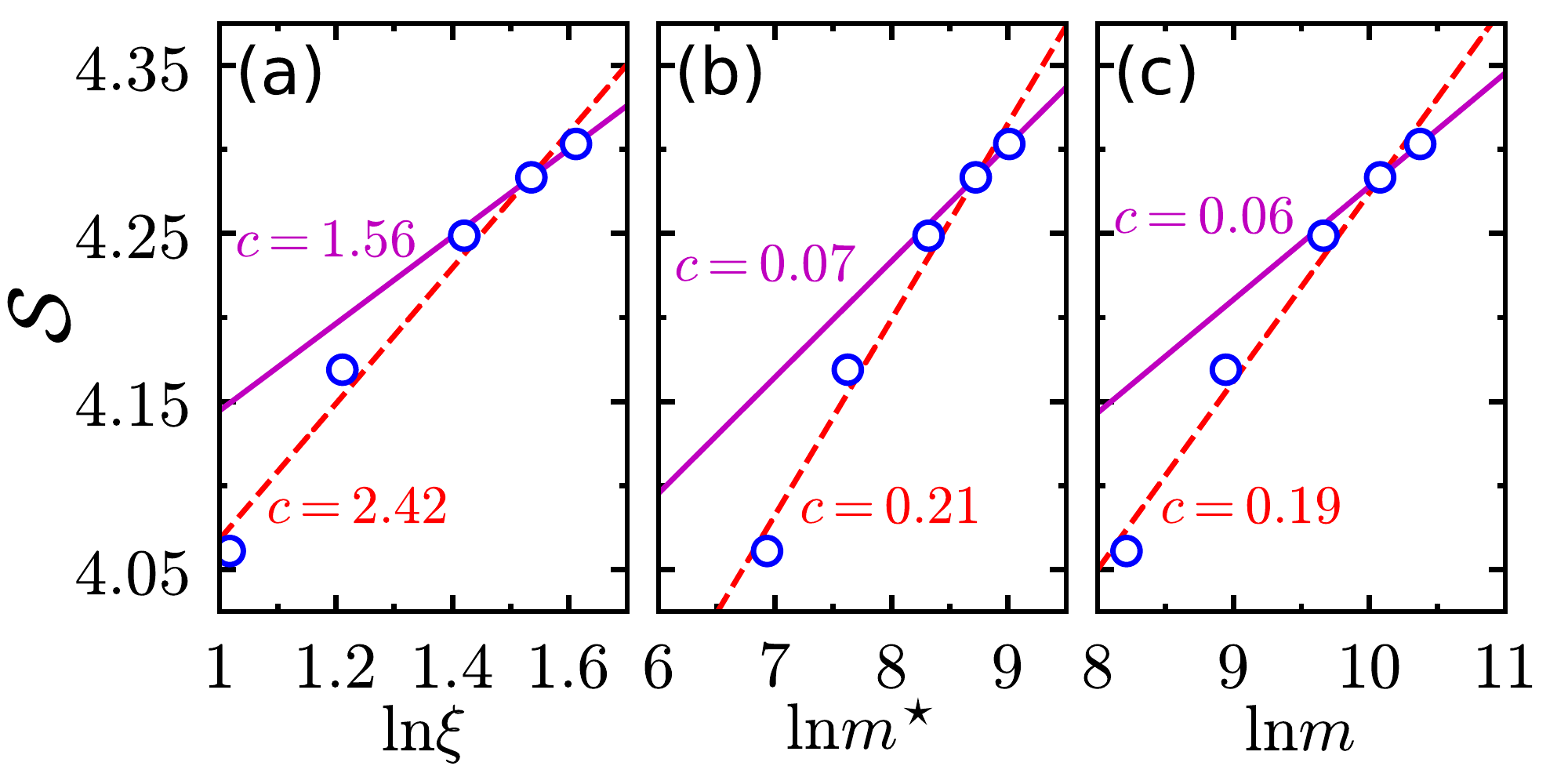}
\caption{Fitting the central charge of the even sector state on the YC$8$-$0$ cylinder.
We use $SU(2)$-infinite-DMRG and central charge is obtained by fitting entanglement entropy with (a) the largest correlation length $\ln\xi$, ${\cal S} = (c/6) \ln\xi + a$; (b) number of $SU(2)$ bases $m^{\star}$, ${\cal S} =  1/(\sqrt{12/c}+1)\ln m^{\star} + b^{\star}$; and (c) number of $U(1)$ bases $m$, ${\cal S} =  1/(\sqrt{12/c}+1)\ln m + b$. 
The five data points correspond to $SU(2)$ bond dimensions $m^{\star}=1024$, $2048$, $4096$, $6144$ and $8192$. 
Maximal equivalent-$U(1)$ bond dimension $m=32014$.
Fitting with five points leads to red dashed lines while fitting with two largest points gives us magenta solid ones.
\label{fig:entcorrlen}
} 
\end{figure}

We remark that a parallel work~\cite{Shoushu2019} reported a large central charge $c=5$ for the even sector state on a finite YC$8$-$0$ cylinder with a small length $L_x$.
However, our data of the even sector state on YC$8$-$0$ cylinder does not agree with their observation.
We use $SU(2)$-infinite-DMRG and have simulated bond dimension up to $m^{\star}=8192$ (equivalent to $U(1)$-DMRG with $m=32014$).
In Fig.~\ref{fig:entcorrlen}, we fit the central using three different methods~\cite{Pollmann_2009}, namely the scaling of entropy with (a) the correlation length, (b) the $SU(2)$-DMRG bond dimension, and (c) the equivalent $U(1)$-DMRG bond dimension. 
The fitted central charge is much smaller than $c=5$, and it keeps on decreasing as bond dimension increases.
The central charge is likely to eventually go to $c=0$ at the infinite bond dimension.
So the observation of $c=5$ in Ref.~\cite{Shoushu2019} might be a finite size effect: They only simulated a small cylinder ($L_x=16$), but the central charge is only well defined in the 1D limit with $L_x\gg L_y =8$.

We also note that there is a weak ($C_6$ breaking) nematic ordering in the spin liquid groundstates, namely a small difference of bond strengths in three inequal directions. 
The cylindrical geometry of DMRG breaks $C_6$ symmetry explicitly, so it is not surprising that a gapless $U(1)$ Dirac spin liquid weakly breaks $C_6$~\cite{He2018}. 
Second, the $C_6$ breaking is weak and it becomes even weaker as we increase the bond dimension (Tab.~\ref{tab:nematicity}). The YC$6$-$0$ cylinder has extremely tiny nematicity, which may come from the fact that DMRG simulation is fully converged for such a small $L_y$.

\begin{table}[!tbh]
\caption{\label{tab:nematicity} Discrepancy of bond strengths (subtracted by the average value $\sim-0.18$) in the lowest-energy odd-sector state for various YC$L_y$-$0$ cylinders by setting $J_2/J_1=0.12$ and $\theta=0$. They are shown in three inequal directions $\vec{a}_2$, $\vec{a}_1$ and $\vec{a}_1 - \vec{a}_2$. The data is accurate to $4$ decimal places.}
\setlength{\tabcolsep}{0.25cm}
\renewcommand{\arraystretch}{1.4}
\begin{tabular}{ccccc}
\hline\hline
 Geometry & $m$ & $\vec{a}_2$ & $\vec{a}_1$ & $\vec{a}_1 - \vec{a}_2$ \\ \hline
YC$6$-$0$ & $1024$ & \ \ $0.0002$ & $-0.0004$ & \ \ $0.0002$ \\
 & $2048$ & $-0.0004$ & \ \ $0.0007$ & $-0.0004$ \\
 & $4096$ & $-0.0005$ & \ \ $0.0011$ & $-0.0005$ \\
  & $6144$ & $-0.0006$ & \ \ $0.0012$ & $-0.0006$ \\ \hline
YC$8$-$0$ & $1024$ & \ \ $0.0286$ & $-0.0576$ & \ \ $0.0290$ \\
 & $2048$ & \ \ $0.0251$ & $-0.0503$ & \ \ $0.0252$ \\
 & $4096$ & \ \ $0.0227$ & $-0.0455$ & \ \ $0.0227$ \\
  & $6144$ & \ \ $0.0218$ & $-0.0437$ & \ \ $0.0219$ \\
  & $8192$ & \ \ $0.0214$ & $-0.0428$ & \ \ $0.0214$ \\
  & $12288$ & \ \ $0.0210$ & $-0.0419$ & \ \ $0.0210$ \\ \hline
 YC$10$-$0$ & $1024$ & \ \ $0.0517$ & $-0.1072$ & \ \ $0.0555$ \\
 & $2048$ & \ \ $0.0459$ & $-0.0927$ & \ \ $0.0468$ \\
 & $4096$ & \ \ $0.0404$ & $-0.0818$ & \ \ $0.0413$ \\
 & $6144$ & \ \ $0.0381$ & $-0.0765$ & \ \ $0.0384$ \\
 & $8192$ & \ \ $0.0365$ & $-0.0732$ & \ \ $0.0368$ \\ \hline
YC$12$-$0$ & $1024$ & \ \ $0.0682$ & $-0.1481$ & \ \ $0.0798$ \\
 & $2048$ & \ \ $0.0623$ & $-0.1319$ & \ \ $0.0696$ \\
 & $4096$ & \ \ $0.0573$ & $-0.1171$ & \ \ $0.0598$ \\
 & $6144$ & \ \ $0.0534$ & $-0.1093$ & \ \ $0.0559$ \\
 & $8192$ & \ \ $0.0509$ & $-0.1042$ & \ \ $0.0533$ \\
\hline \hline
\end{tabular}
\end{table}

\section{Additional numerical data}

\subsection{Correlation length spectrum}

We have in total simulated $16$ different geometries and/or system sizes.
Similar to the $4$ clusters shown in main text, the fermion bilinear and monopole operators also show up in correlation length spectrum of all the other $12$ clusters. 
Here provide these data in this appendix.

\begin{figure}[b]
\includegraphics[width=0.8\linewidth]{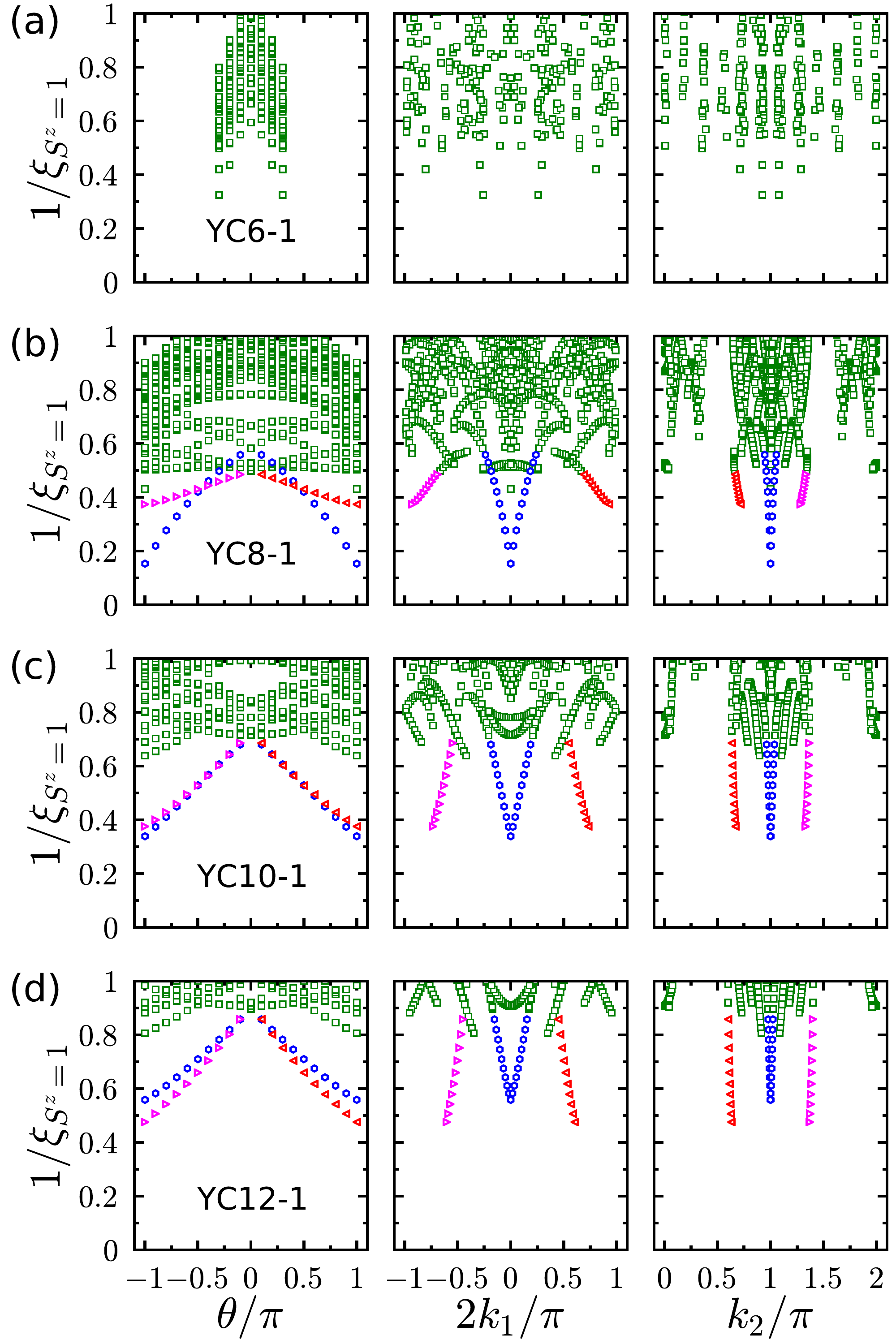}
\caption{Inverse correlation length $1/\xi_{S^z=1}$ (green  {\scriptsize $\square$}) as a function of the flux $\theta$ (left column), momentum $2k_1/\pi$ (middle column) and momentum $k_2/\pi$ (right column) for YC$L_{y}$-$1$ cylinders by setting $J_2/J_1=0.12$. We choose bond dimension (a) $m=4096$ for $L_{y}=6$, (b) $12288$ for $L_{y}=8$, (c) $12288$ for $L_{y}=10$ and (d) $12288$ for $L_{y}=12$ respectively. Specially, we denote the lowest-lying spinon-pair excitations by blue $\varhexagon$ and monopole excitations by red $\triangleleft$ and magenta $\triangleright$.
\label{fig:specs2}
} 
\end{figure}

\subsubsection{YC$2n$-$(2k+1)$}

\begin{figure}[b]
  \includegraphics[width=0.8\linewidth]{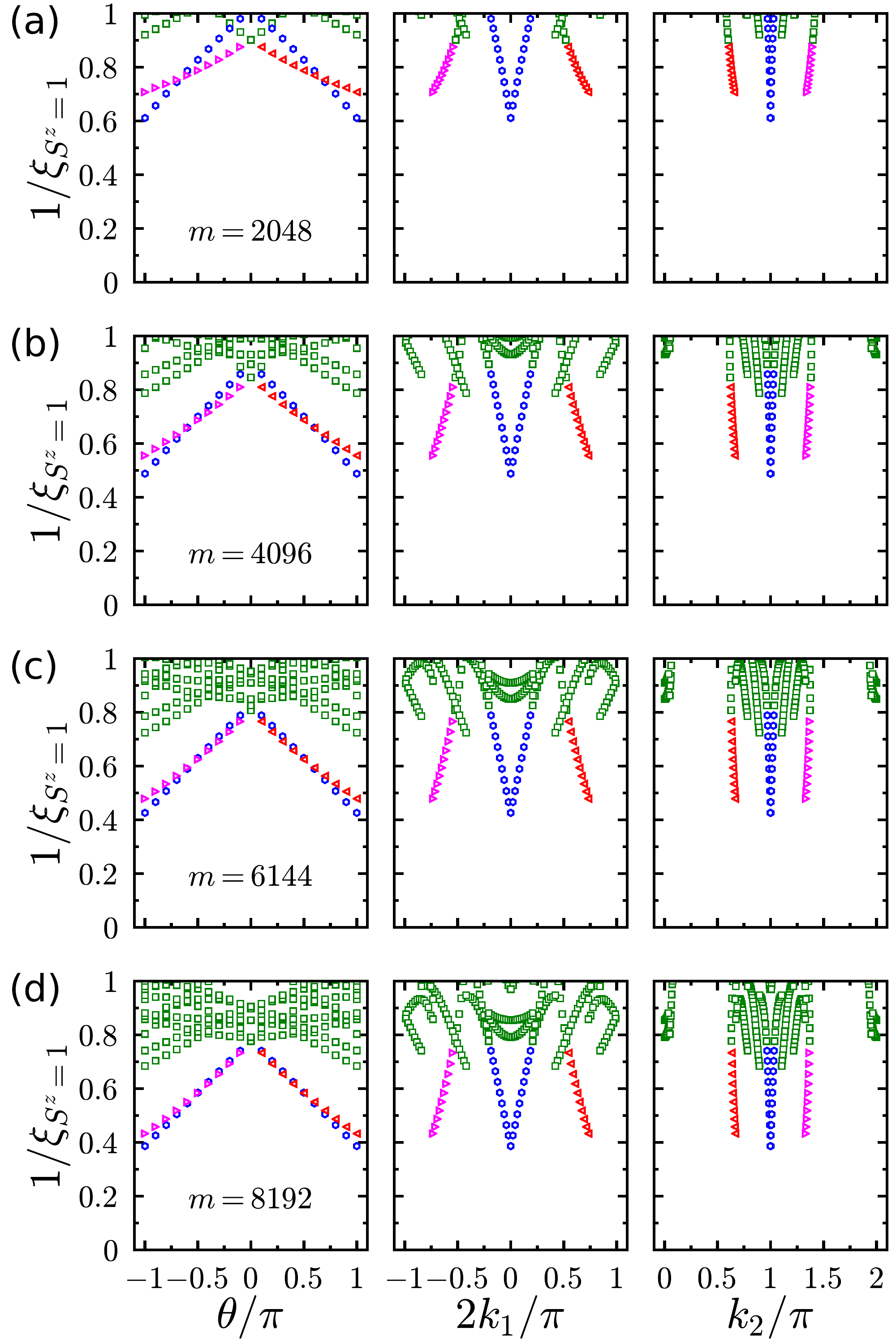}
  \caption{Inverse correlation length $1/\xi_{S^z=1}$ (green {\scriptsize $\square$}) as a function of the flux $\theta$ (left column), momentum $2k_1/\pi$ (middle column) and momentum $k_2/\pi$ (right column) for YC$10$-$1$ cylinders by setting $J_2/J_1=0.12$. Bond dimension (a) $m=2048$, (b) $4096$, (c) $6144$ and (d) $8192$ respectively. Specially, we denote the lowest-lying spinon-pair excitations by blue $\varhexagon$ and monopole excitations by red $\triangleleft$ and magenta $\triangleright$.
\label{fig:specs1}
} 
\end{figure}

Let us first look at the YC$2n$-$(2k+1)$ cylinder.
We have already shown results of YC$8$-$1$ and YC$10$-$1$ cylinders in the main text, but for comparison we still plot them together with YC$6$-$1$ and YC$12$-$1$ ones in Fig.~\ref{fig:specs2}. 
For YC$6$-$1$ in the panel (a), we can insert the flux adiabatically until the flux $\theta>0.3\pi$. 
Once a Dirac spin liquid is put on a small cylinder, it might have an instability by spontaneously generating massive terms during flux insertion. 
Therefore sometimes the numerical adiabatic change breaks down, but such finite size effect will be gone in the pure $2$+$1$D limit, so only data is shown where the ground state remains in the spin liquid state. 
This is consistent with our observation that, for a larger system size (i.e. YC$8$-$1$, YC$10$-$1$ and YC$12$-$1$), the adiabatic twist can be maintained even when $\theta=\pi$. 
The trend of the spectrum is similar to larger system sizes, although the information of $\theta\in(0,0.3\pi)$ is not enough for us to mark the type of its excitations. 
For other three cylinders, we also find one Dirac cone (low-lying excitations) is at a $M$ point $(2k_1,k_2)=(0,\pi)$ ($M_2$ or $M_3$ point) and the other is close to the $K_{\pm}$ point $(2k_1,k_2)=\pm(2\pi/3,2\pi/3)$.
The former correspond to Fermion bilinears and the latter correspond to monopole operators.

Comparing the correlation length spectrum of various $L_y=8$, $10$ and $12$, it seems that the larger the system is, the higher the Dirac mode (blue $\varhexagon$) is. 
We think it is an artifact from the finite entanglement effect (bond dimensions) in infinite-DMRG simulations. 
For larger system sizes ($L_y$), the gapless modes suffers more severe truncation error, yielding a smaller correlation length at a given bond dimension.
This artifact can be seen in Fig.~\ref{fig:specs1}, where clearly the Dirac mode becomes lower as the bond dimension increases. 
On the other hand, to achieve the same accuracy (truncation error), the required bond dimension increases exponentially with the circumference of the cylinder. Table~\ref{tab:trunerr} shows the truncation error for different bond dimensions, system sizes and flux. 
For the large system size (YC$10$-$1$, YC$12$-$1$), $m = 12288$ roughly gives comparable accuracy as $m = 1024$ for YC$8$-$1$. 
Therefore, more care should be taken if one wants to compare the results between different system sizes.

\begin{figure}[b]
  \includegraphics[width=0.8\linewidth]{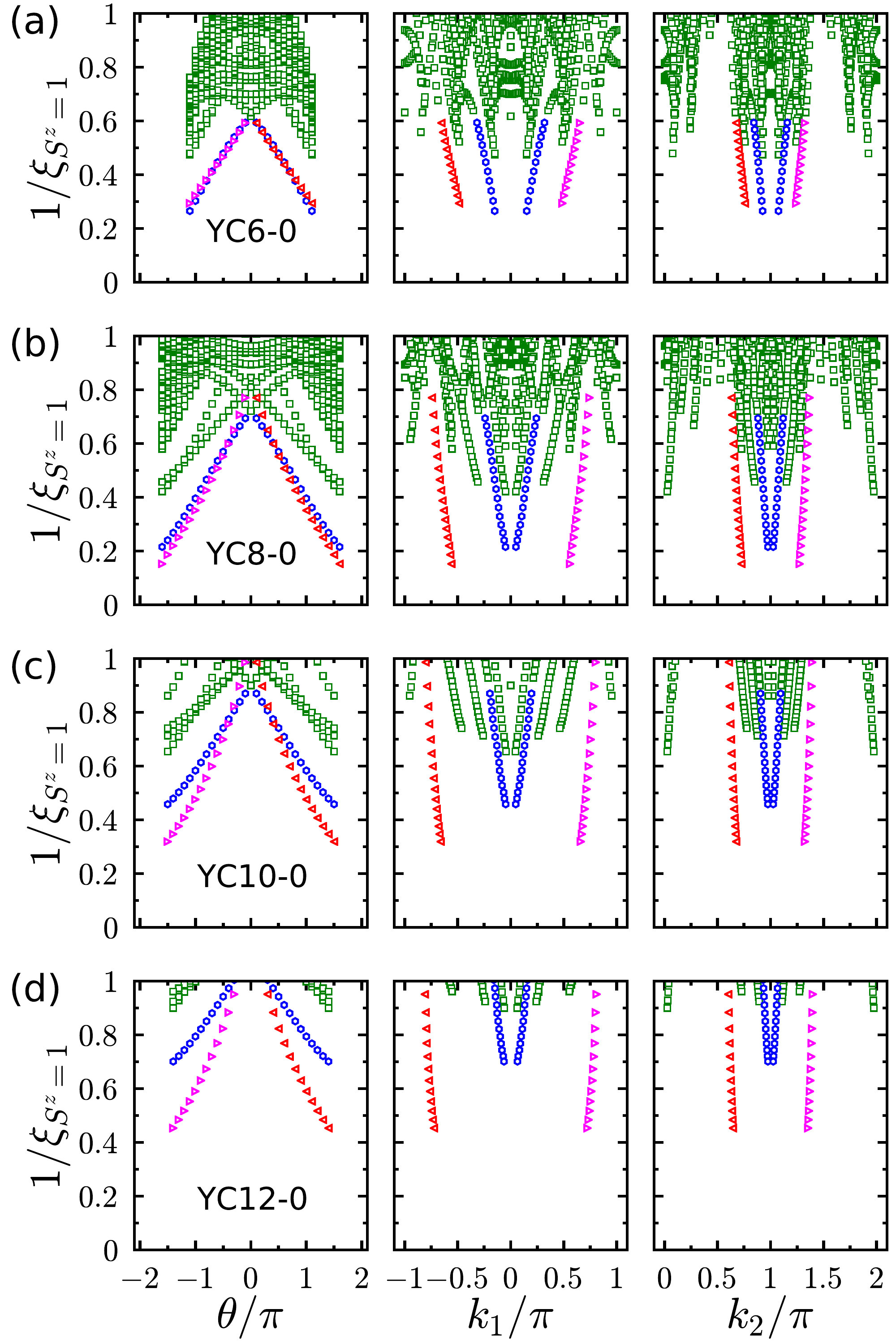}
  \caption{Inverse of correlation length $1/\xi_{S^z=1}$ (green {\scriptsize $\square$}) as a function of the flux $\theta$ (left column), momentum $k_1/\pi$ (middle column) and momentum $k_2/\pi$ (right column) for (a) YC$6$-$0$, (b) YC$8$-$0$, (c) YC$10$-$0$, and (d) YC$12$-$0$ cylinders by setting $J_2/J_1=0.12$. Bond dimension $m=6144$ for all cases. Specially, we denote the lowest-lying spinon-pair excitations by blue $\varhexagon$, while monopole excitations by red $\triangleleft$ and magenta $\triangleright$.
 The data is shown for the lowest energy ``topological" sector, namely odd sector (sometimes called spinon sector)~\cite{McCulloch2016}.
\label{fig:specs3}
} 
\end{figure}

\begin{figure}[b]
  \includegraphics[width=0.8\linewidth]{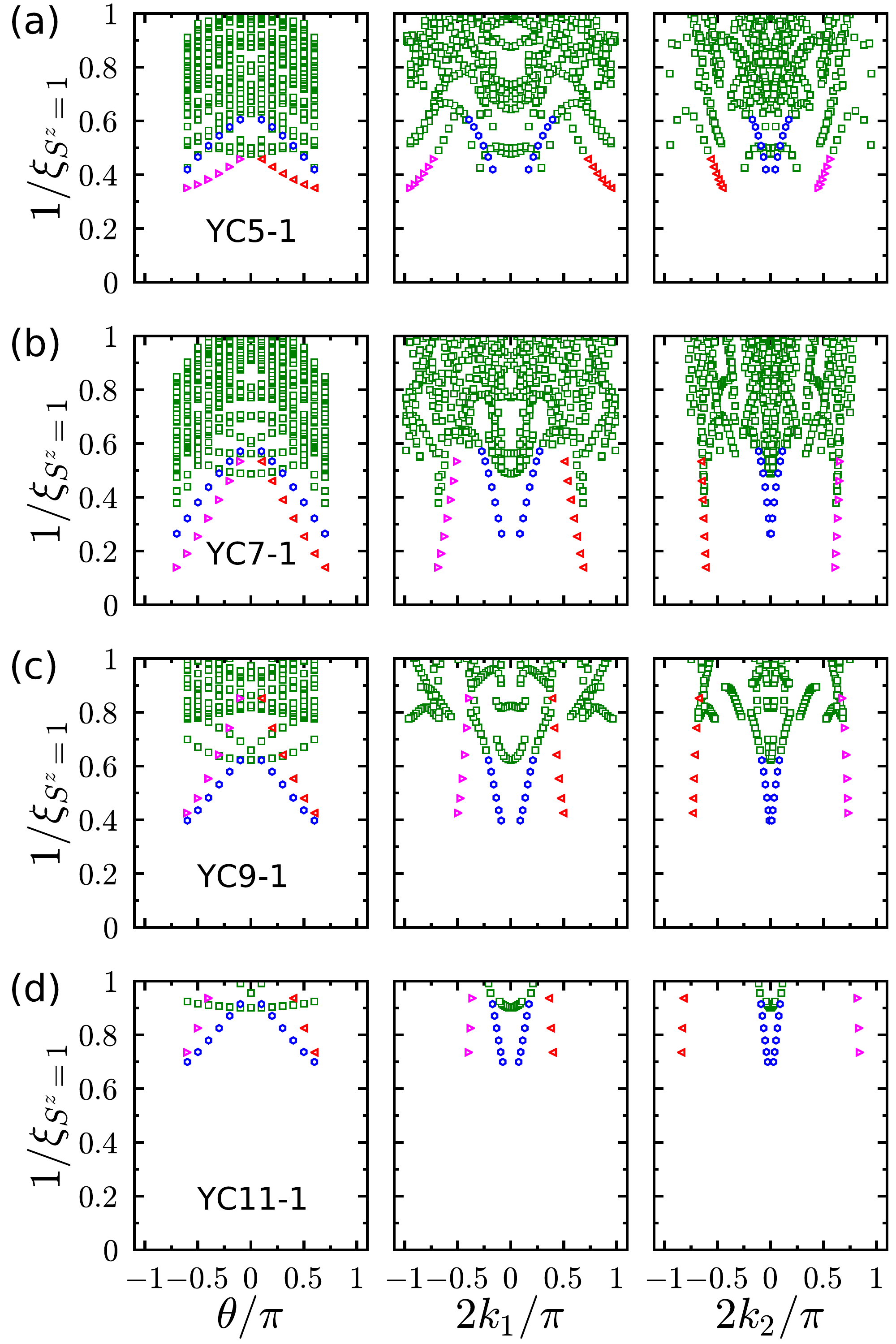}
  \caption{Inverse of correlation length $1/\xi_{S^z=1}$ (green {\scriptsize $\square$}) as a function of the flux $\theta$ (left column), momentum $2k_1/\pi$ (middle column) and momentum $2k_2/\pi$ (right column) for YC$L_{y}$-$1$ cylinders by setting $J_2/J_1=0.12$. Bond dimension (a) $m=4096$ for $L_{y}=5$, (b) $8192$ for $L_{y}=7$, (c) $6144$ for $L_{y}=9$, and (d) $4096$ for $L_{y}=11$ respectively. Specially, we denote the lowest-lying spinon-pair excitations by blue $\varhexagon$, while monopole excitations by red $\triangleleft$ and magenta $\triangleright$.
\label{fig:specs4}
} 
\end{figure}

\begin{table}[!tbh]
\caption{\label{tab:trunerr} Truncation error for various YC$2n$-$1$ cylinder by setting $J_2/J_1=0.12$. $\theta=\pi$ has much bigger truncation error than $\theta=0$.}
\setlength{\tabcolsep}{0.25cm}
\renewcommand{\arraystretch}{1.4}
\begin{tabular}{ccccc}
\hline\hline
                    & $m=1024$ & $m=6144$ & $m=12288$ \\ \hline
YC$8$-$1$\ \ \ $(\theta=0)$ & $5.3\times10^{-5}$ & $5.3\times10^{-6}$ & $1.7\times10^{-6}$  \\
YC$8$-$1$\ \ \ $(\theta=\pi)$ & $6.5\times10^{-5}$ & $1.1\times10^{-5}$ & $4.5\times10^{-6}$ \\ \hline
YC$10$-$1$ $(\theta=0)$ & $1.3\times10^{-4}$ & $2.5\times10^{-5}$ & $1.3\times10^{-5}$  \\
YC$10$-$1$ $(\theta=\pi)$ & $1.5\times10^{-4}$ & $3.7\times10^{-5}$ & $2.1\times10^{-5}$ \\ \hline
YC$12$-$1$ $(\theta=0)$ &  & $5.8\times10^{-5}$ & $3.4\times10^{-5}$  \\
YC$12$-$1$ $(\theta=\pi)$ &  & $7.1\times10^{-5}$ & $4.4\times10^{-5}$ \\
\hline \hline
\end{tabular}
\end{table}

\subsubsection{YC$2n$-$2k$}

\begin{figure}[t]
  \includegraphics[width=0.8\linewidth]{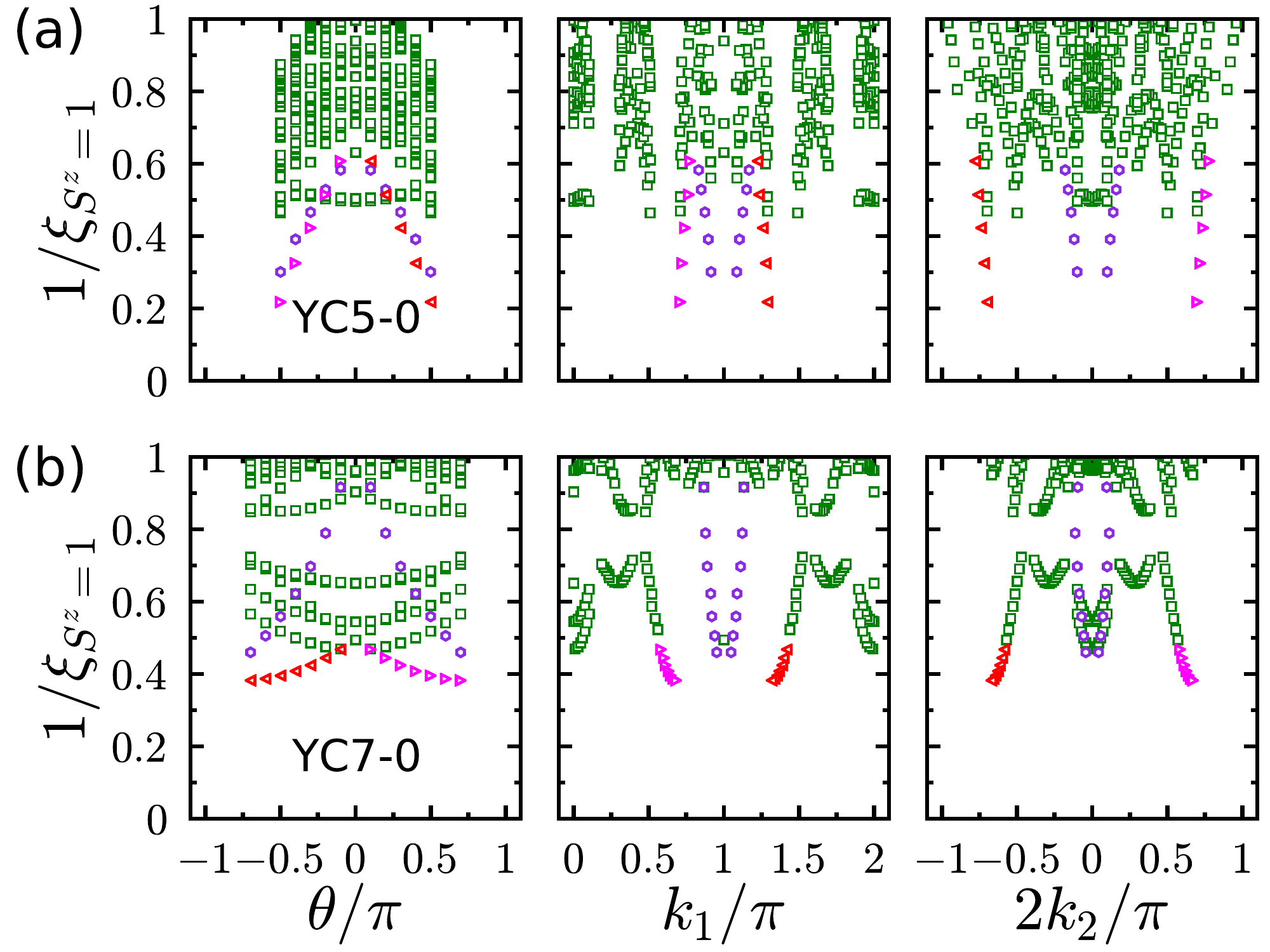}
  \caption{Inverse of correlation length $1/\xi_{S^z=1}$ (green {\scriptsize $\square$}) as a function of the flux $\theta$ (left column), momentum $k_1/\pi$ (middle column) and momentum $2k_2/\pi$ (right column) for (a) YC$5$-$0$ and (b) YC$7$-$0$ cylinders by setting $J_2/J_1=0.12$. Bond dimension $m=2048$ for both cases. Specially, we denote the lowest-lying spinon-pair excitations by violet $\varhexagon$, while monopole excitations by red $\triangleleft$ and magenta $\triangleright$.
\label{fig:specs5}
} 
\end{figure}

\begin{figure}[b]
  \includegraphics[width=0.8\linewidth]{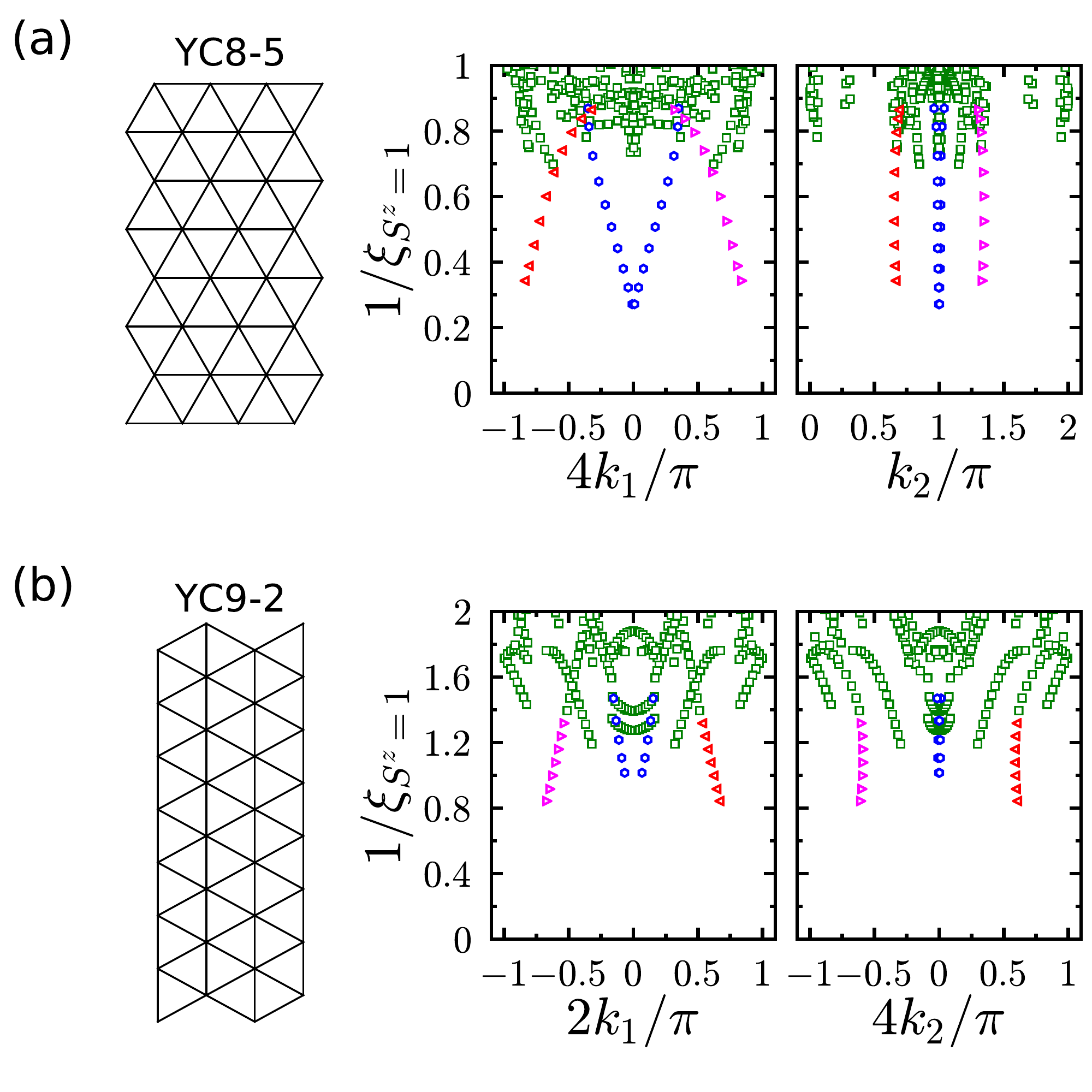}
  \caption{Inverse of correlation length $1/\xi_{S^z=1}$ (green {\scriptsize $\square$}) as a function of the momentum $k_1/\pi$ (middle column) and momentum $k_2/\pi$ (right column) for geometry (a)YC$8$-$5$ (e.g. ``XC"$8$-$1$~\cite{ZZhu2015}) and (b)YC$9$-$2$ (``YC"$8$-$2$~\cite{Wenjun2015}) by setting $J_2/J_1=0.12$. Bond dimension $m=4096$ for both cases. Specially, we denote the lowest-lying spinon-pair excitations by blue $\varhexagon$, while monopole excitations by red $\triangleleft$ and magenta $\triangleright$.
\label{fig:specs6}
} 
\end{figure}

Next we look at the YC$2n$-$2k$ cylinder. As we discussed previously, this class of cylinder behaves very different than YC$2n$-$(2k+1)$ discussed above. For the $U(1)$ Dirac spin liquid, the YC$2n$-$(2k+1)$ cylinder hits the gapless Dirac cone at $\theta=\pi$, while YC$2n$-$2k$ cylinder hits the gapless Dirac cone at $\theta=2\pi$. Our simulation on the YC$2n$-$2k$ cylinder is also consistent with this scenario. For YC$6$-$0$, the adiabaticity of the twist can be maintained until $\theta=1.1\pi$, after which the system collapses to the other topological sector. For YC$8$-$0$, YC$10$-$0$ and YC$12$-$0$ cylinders, adiabatic twist can persist until $\theta\approx1.5\pi$. 

Fig.~\ref{fig:specs3} shows the $S^z=1$ correlation length spectrum of the YC$6$-$0$, YC$8$-$0$, YC$10$-$0$ and YC$12$-$0$ cylinders at $J_2/J_1 = 0.12$. 
We find the lowest modes behave like the fermion bilinears and monopole operators of $U(1)$ DSL.
Similar to the YC$2n$-$(2k+1)$ cylinder, the lowest modes show a linear dependence with the flux $\theta$.
The lowest-lying fermion bilinear appears at $M_2 = (0,\pi)$ point (labeled by $(k_1,k_2)$). 
We do not find spinon-pair excitations at $M_1$ and $M_3$, which can be understood by the explicitly broken rotation symmetry $C_6$ on a finite-width and infinite-length cylinder. 
Importantly, we also find that one additional branch appears nearby $K_{\pm}$ points $(k_1,k_2)=\pm(-2\pi/3,2\pi/3)$, signaling the expected monopole excitations.

\begin{figure}[t]
  \includegraphics[width=0.8\linewidth]{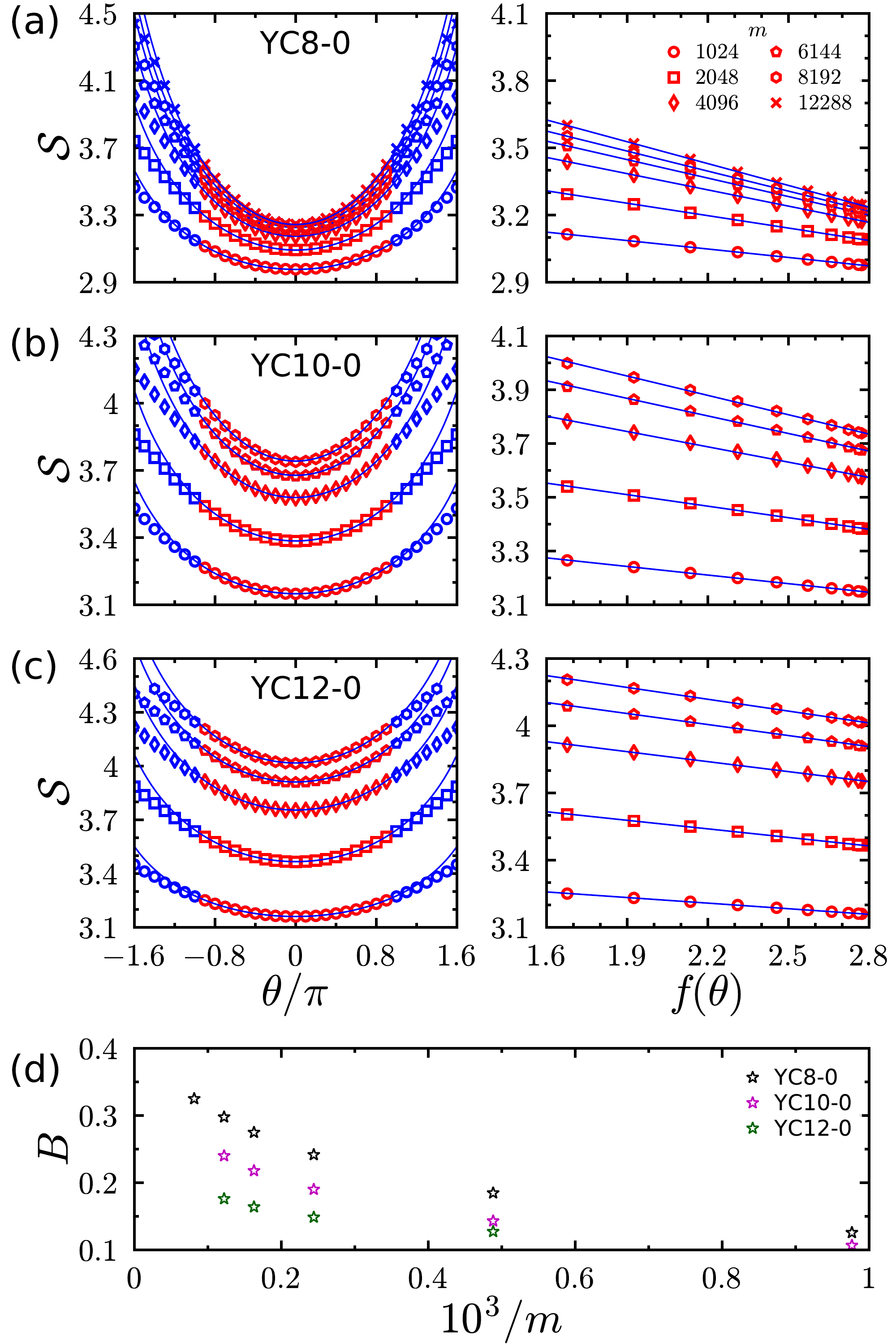}
  \caption{Entanglement entropy $\cal S$ as a function of the flux angle $\theta$ (left column) and $f(\theta)=\sum_{n=1}^{N_f} \ln \left|2\sin \left[s\left(\theta-\theta^c_n\right) / 2\right] \right|$ (right column) for various cylinders: (a) YC$8$-$0$, (b) YC$10$-$0$ and (c) YC$12$-$0$. Bond dimensions $m=2048$ ({\scriptsize $\square$}), $4096$ ({\scriptsize $\diamondsuit$}), $6144$ ($\pentagon$), $8192$ ($\varhexagon$) and $12288$ ($\times$). We do the fitting of data (red symbols) around minima to the Eq.~(2). The best fitting (blue solid line) give us the coefficient $B$ which is marked in the panel (d) as a function of $m$.
\label{fig:scalings7}
} 
\end{figure}

\subsubsection{YC$(2n+1)$-$(2k+1)$}

Thirdly, the YC$(2n+1)$-$(2k+1)$ cylinder is basically the same as YC$2n$-$(2k+1)$. For the YC$(2n+1)$-$(2k+1)$ cylinder, we expect that spinons hit Dirac cones when $\theta=\pi$ or $3\pi$ independent of the emergent gauge flux $\phi=0$ or $\pi$. 
In Fig.~\ref{fig:specs4}, we plot the $S^z=1$ correlation length spectrum as a function of the flux $\theta$, $2k_1$ and $2k_2$ respectively. Different from YC$2n$-$(2k+1)$ cylinder, the adiabatic twist cannot persist to $\pi$ due to the small gap and instability of the state as Dirac cone is approached. 
Furthermore, from the momentum-resolved spectrum we find Fermion bilinears appearing at a $M$ point $(2k_1, 2k_2)=(0,0)$, and monopoles close to the $K_{\pm}$ points $(2k_1, 2k_2)=\pm(2\pi/3,-2\pi/3)$.

\begin{figure}[t]
  \includegraphics[width=0.8\linewidth]{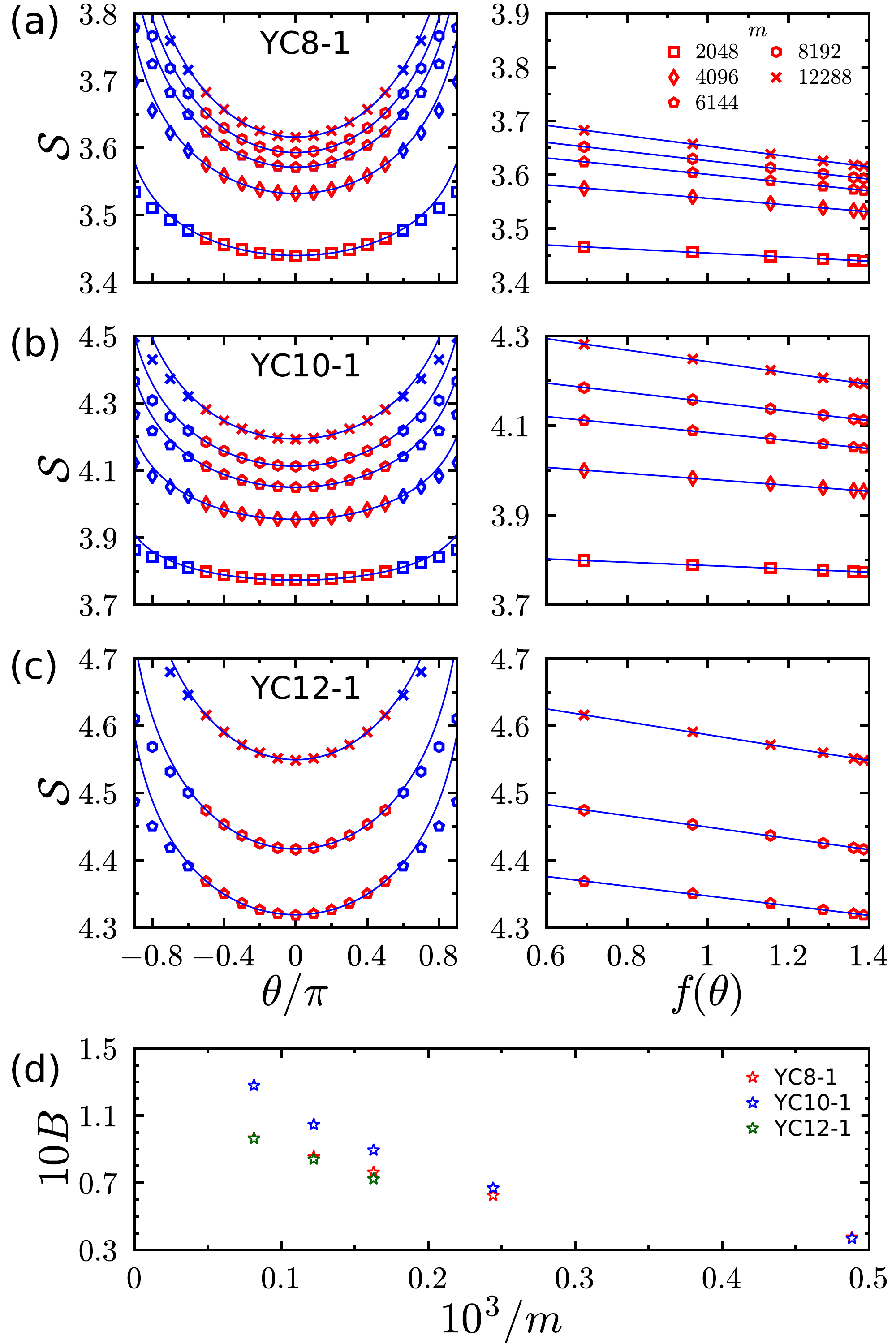}
  \caption{Entanglement entropy $\cal S$ as a function of the flux angle $\theta$ (left column) and $f(\theta)=\sum_{n=1}^{N_f} \ln \left|2\sin \left[s\left(\theta-\theta^c_n\right) / 2\right] \right|$ (right column) for various cylinders: (a) YC$8$-$1$, (b) YC$10$-$1$ and (c) YC$12$-$1$. Bond dimensions $m=1024$ ($\Circle$), $2048$ ({\scriptsize $\square$}), $4096$ ({\scriptsize $\diamondsuit$}), $6144$ ($\pentagon$), $8192$ ($\varhexagon$) and $12288$ ($\times$). We do the fitting of data (red symbols) around minima to the Eq.~(2). The best fitting (blue solid line) give us the coefficient $B$ which is marked in the panel (d) as a function of $m$.
\label{fig:scalings8}
} 
\end{figure}

\subsubsection{YC$(2n+1)$-$2k$}

In Fig.~\ref{fig:specs5}, we plot the $S^z=1$ correlation length spectrum of YC$(2n+1)$-$0$ cylinders as a function of the flux $\theta$, $k_1$ and $2k_2$ respectively. We find that the spinon-pair excitations appear at $(k_1,2k_2)=(\pi,0)$ ($M_1$ or $M_3$ point). Additionally, monopole excitations appears nearby $K_{\pm}$ $(k_1,2k_2)=\pm(-2\pi/3,-2\pi/3)$.

\subsubsection{Other geometries}

In previous studies, people use different geometries, such as ``XC" geometry~as defined in Ref.~\cite{ZZhu2015} (left column of Fig.~\ref{fig:specs6}~(a)) and ``YC" geometry~as defined in Ref.~\cite{Wenjun2015} (left column of Fig.~\ref{fig:specs6}~(b)). Therefore, we also analyze the adiabatic flux insertion for cylinders ``XC"$8$-$1$ and ``YC"$8$-$2$ as shown in Fig.~\ref{fig:specs6}.

For the ``XC"$8$-$1$ cylinder (equivalent to the YC$8$-$5$ cylinder), the spinon-pair excitations hit Dirac cones when $\theta=\pi$ independent of the emergent gauge flux $\phi=0$ or $\pi$. In Fig.~\ref{fig:specs6}~(a), we find that its Dirac mode appears at a $M$ point $(4k_1,k_2)=(0,\pi)$ modulo $2\pi$. Monopole excitations are very close to $K_{\pm}$ points $(4k_1,k_2)=\pm(-2\pi/3,2\pi/3)$.

``YC" cylinders can be transformed to ``XC" by a rotation $\pi/2$ in the $xy$-plane. For the ``YC"$8$-$2$ cylinder (equivalent to the YC$9$-$2$ cylinder), the spinon-pair excitations hit a Dirac cone when $\theta=\pi$ independent of the emergent gauge flux $\phi=0$ or $\pi$ too. In Fig.~\ref{fig:specs6}~(b), we find that its Dirac mode appears at a $M$ point $(2k_1,4k_2)=(0,0)$. Monopole excitations are very close to $K_{\pm}$ points $(2k_1,4k_2)=\pm(2\pi/3,2\pi/3)$ too.

\subsection{Scaling behavior of entanglement entropy}

In addition to Fig.~4, we show more data of entanglement entropy $\cal S$ for the cylinder YC$L_y$-$0$ (Fig.~\ref{fig:scalings7}) and YC$L_y$-$1$ (Fig.~\ref{fig:scalings8}). Firstly, we find that Eq.~(2) accurately fits the data around the minimal value of $\cal S$ for all the geometries and bond dimensions. Secondly, we notice that the fitting parameter $B$ shows strong dependence on bond dimension, system geometry, etc.. 
The strong dependence on bond dimension makes it hard to draw a conclusion on the question whether $B$ is a universal quantity or not. However, it is clear that the entanglement entropy always follows the universal scaling law conjectured for the $U(1)$ DSL.\\

\subsection{Gaps}

\subsubsection{Gap measurement}
\begin{figure}[h]
  \includegraphics[width=0.8\linewidth]{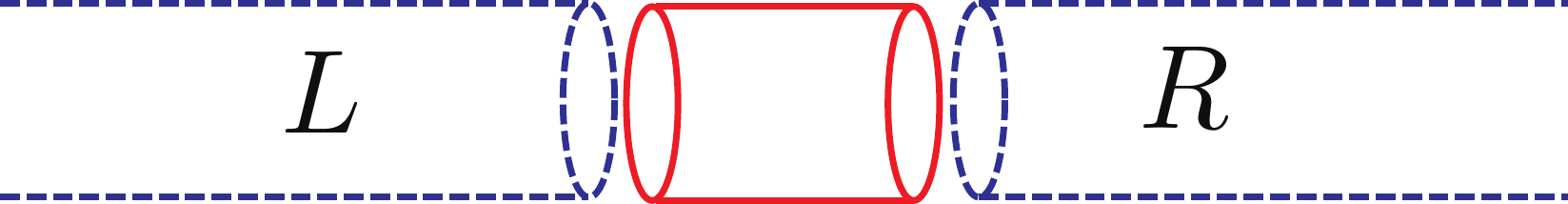}
  \caption{Schematic picture for the gap measurement in infinite-DMRG.
\label{fig:gaps9}
} 
\end{figure}

We use an algorithm that combines infinite-DMRG and finite DMRG to calculate the spin gap in Fig.~\ref{fig:gaps9}. We first obtain a converged wave-function of an infinitely-long cylinder using infinite-DMRG ``warm-up" steps. Then we insert a sector of cylinder consisting of $L_y\times L_y$ sites (red cylinder) into the middle of two half-chains. The left (L) and right (R) semi-infinite cylinder can be considered as environment (boundary conditions). We further do sweeps inside the small cylinder and get the lowest energy $E_0 (S^z = 1)$ in the $S^z=1$ sector and the energy of the $1^{\text{st}}$ excited state $E_1 (S^z = 0)$ in the $S^z=0$ sector. Finally, we obtain the gap $\Delta_{S^z=1}=E_0(S^z=1) - L^2_y e_0(S^z=0)$ and $\Delta_{S^z=0}=E_1(S^z=0) - L^2_y e_0(S^z=0)$ where we get the average energy per-site $e_0(S^z=0)$ during infinite-DMRG calculations.

\subsubsection{Gap scaling}

\begin{figure}[h]
  \includegraphics[width=0.8\linewidth]{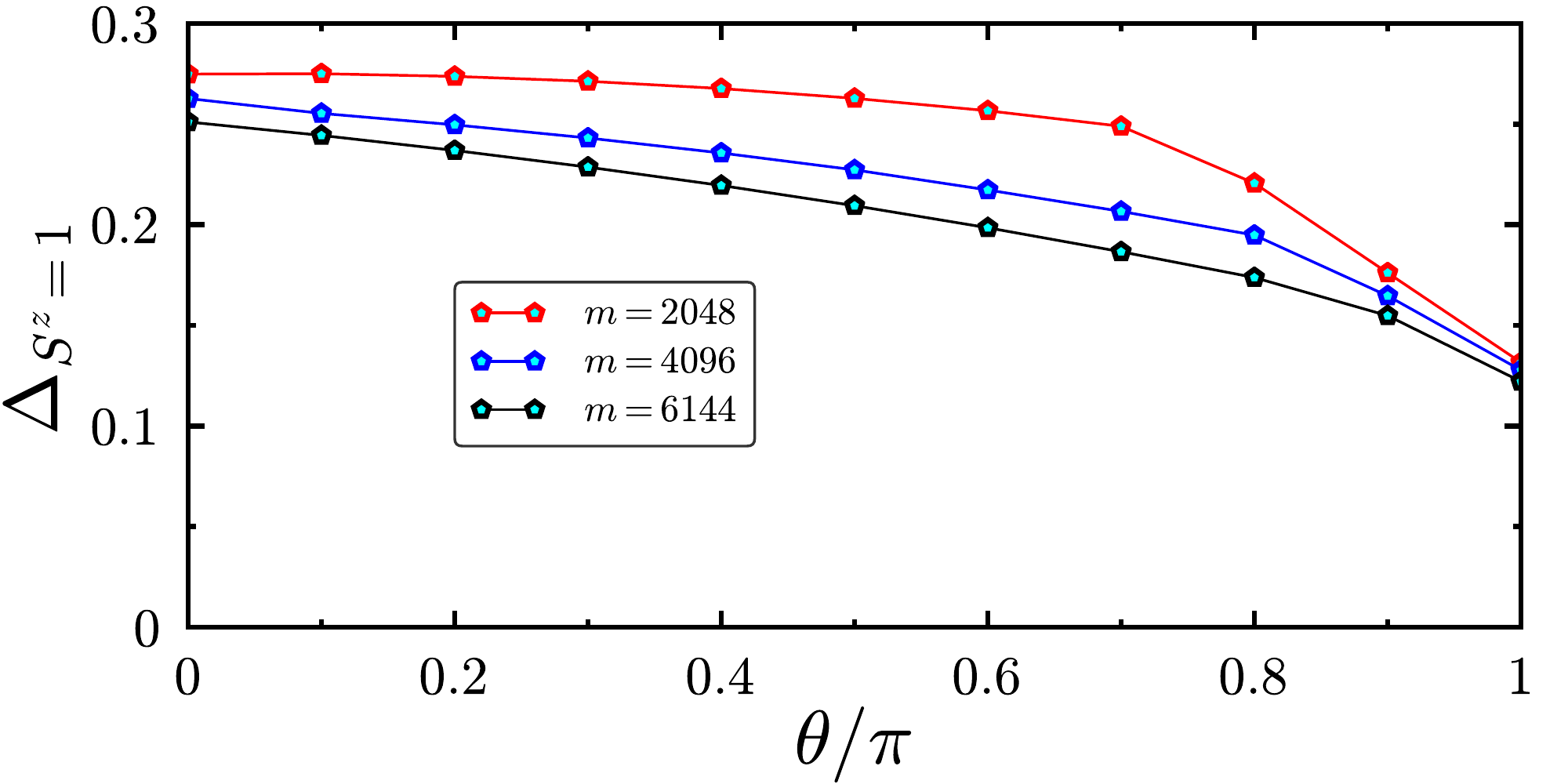}
  \caption{Dependence of the spin gap $\Delta_{S^z=1}$ (solid line) on the spin flux $\theta$ and the truncated bond dimension m. Data is taken with $J_2/J_1=0.12$ on the YC$10$-$1$ cylinder. Generally the estimated gap decreases with the bond dimension $m$: For larger the system sizes, the energy scale from truncation becomes relevant, thereby overestimating the spin gap.
\label{fig:gaps10}
} 
\end{figure}

In addition to Fig.~2, we show more data for the gap in Fig.~\ref{fig:gaps10}.  For large enough $L_y$, the spin gap of a DSL at $\theta=0$ decreases with the circumference size $L_y$ as $\Delta_{S^z=1}\sim v_{S^z=1}/L_y$, but numerically larger $L_y$ also has a larger trunctation error from finite bond dimension $m$, which tends to overestimate the spin gap in Fig.~2.  A simple finite size scaling is therefore difficult, if $m$ sets a larger energy scale.  We therefore demonstrate that the gap value becomes lower with increasing truncated bond dimension in Fig.~\ref{fig:gaps10}, which is consistent with a vanishing gap in the thermodynamic limit ($L_y\rightarrow \infty$).

\end{document}